\documentclass[sigconf,nonacm]{acmart}
\usepackage{multirow}
\usepackage{colortbl}
\usepackage{makecell}
\usepackage{placeins}
\usepackage{hyphenat}
\usepackage{pifont}
\usepackage{tabularx}
\usepackage{listings}
\usepackage{xcolor}

\definecolor{mygreen}{RGB}{0,110,101}
\definecolor{myamber}{RGB}{232,130,21}
\definecolor{myred}{RGB}{192,0,0}

\AtBeginDocument{}

\begin{document}

\title{Beyond Tag Collision: Cluster-based Memory Management for Tag-based Sanitizers}

\author{Mengfei Xie}\thanks{This paper has been accepted to the 2025 ACM SIGSAC Conference on Computer and Communications Security (CCS'25).}
\affiliation{
  \institution{Wuhan University}
  \department{School of Cyber Science and Engineering}
  \city{}
  \country{}
}
\email{mfxie96@whu.edu.cn}

\author{Yan Lin}
\authornote{Jianming Fu and Yan Lin are corresponding authors}
\affiliation{
  \institution{Jinan University}
  \department{College of Cyber Security}
  \city{}
  \country{}
}
\email{yanlin@jnu.edu.cn}

\author{Hongtao Wu}
\affiliation{
  \institution{Wuhan University}
  \department{School of Cyber Science and Engineering}
  \city{}
  \country{}
}
\email{hongtaowu@whu.edu.cn}

\author{Jianming Fu}
\email{jmfu@whu.edu.cn}
\affiliation{
  \institution{Wuhan University}
  \department{School of Cyber Science and Engineering}
  \city{}
  \country{}
}
\authornotemark[1]

\author{Chenke Luo}
\affiliation{
  \institution{Tulane University}
  \department{Department of Computer Science}
  \city{}
  \country{}
}
\email{cluo6@tulane.edu}

\author{Guojun Peng}
\affiliation{
  \institution{Wuhan University}
  \department{School of Cyber Science and Engineering}
  \city{}
  \country{}
}
\email{guojpeng@whu.edu.cn}

\renewcommand{\shortauthors}{Mengfei Xie et al.}

\begin{abstract}
Tag-based sanitizers attach a small ``key'' to each pointer and a matching ``lock'' tag to its target memory object, enabling runtime verification of pointer-object consistency and helping developers to detect potential memory violations. However, the limited tag encoding space challenges existing studies in assigning distinct tags to memory objects across temporal and spatial dimensions, leading to potential tag collisions. Such limitations reduce the probabilistic protection capabilities of sanitizers and make them vulnerable to sophisticated tag probing attacks.
\par
In this paper, we present ClusterTag, a novel cluster-based memory allocator aimed at simultaneously mitigating tag collisions in both temporal and spatial dimensions. The core design of ClusterTag effectively balances the significant mismatch between tag encoding space and memory objects: it divides memory objects into multiple independent clusters, thereby limiting tag collisions to finite chunks within each cluster. 
To mitigate tag collisions across clusters, we design a cluster-grained heap randomization scheme. This approach introduces random address intervals between clusters and further breaks the entropy limitation of the tag space.
ClusterTag has been implemented as an independent memory allocator that seamlessly integrates with tag-based sanitizers such as HWASan, and maintains comparable performance overhead (within 1\%) at various randomization densities.
Security evaluations on the Juliet dataset indicate that ClusterTag exhibits deterministic results across 500 repeated tests (5,652 reported and 1,530 missed), while the existing three types of tag assignment strategies all exhibit probabilistic false negatives due to tag collisions.
Quantitative analysis across three tag collision distance metrics-minimum, average, and unpredictability-demonstrates that ClusterTag achieves balanced improvements across all three, whereas prior tag assignment schemes (random, staggered, fixed) show significant trade-offs in at least one metric.

\end{abstract}

\ccsdesc[500]{Security and privacy~Software security engineering}

\keywords{Tag Collision, Sanitizer, Memory Corruption Vulnerability}

\maketitle

\section{Introduction}
Benefiting from the widespread adoption of modern 64-bit architectures, tag-based sanitizers embed a compact object identifier in the unused high-order bits of pointers and validate it on every memory access. Unlike location-based solutions such as AddressSanitizer (ASan) \cite{asan}, they do not need to surround memory objects with red zones to capture illegal accesses and therefore typically deliver stronger security, lower overhead, and better backward compatibility. However, the current three mainstream tagged memory mechanisms-TBI (Top-Byte-Ignore) \cite{tbi_hw}, LAM (Linear-Address-Masking) \cite{lam_hw}, and MTE (Memory Tagging Extension) \cite{mte_hw}-only provide 8-bit, 4-bit, and 6-bit tag spaces, respectively. This hardware constraint prevents sanitizers from assigning unique tags for each memory object, causing potential collisions in both spatial and temporal dimensions \cite{mte_analysis1,mte_analysis2}. Moreover, recent research \cite{stickytag,kim2024tiktag} has shown that attackers can exploit speculative execution to probe the tag space in MTE, thereby stealthily searching for memory objects that share the same ``key-lock'' pairs. 
\par
Due to hardware limitations in expanding tag encoding space, existing research \cite{hwasan,scudo,stickytag} primarily focuses on optimizing tag assignment strategies in memory allocators to mitigate tag collisions. This task involves complex multi-objective optimization challenges, specifically calculating collision distances across both adjacent tags (spatial dimension) and historical tags (temporal dimension) for all memory objects. Intuitively, computing optimal tags in real time is impractical for memory allocators with thousands of allocations per second. As a result, existing solutions primarily employ three types of lightweight assignment algorithms: random \cite{hwasan,ptmalloc}, staggered \cite{scudo}, and fixed assignment \cite{partition,stickytag}, each algorithm provides different probabilistic protection against various vulnerability types. We systematically evaluate these algorithms using three key metrics: minimum collision distance (to prevent attacks on adjacent objects), average collision distance (to reduce overall collision probability), and unpredictability (to ensure entropy against brute-force attacks). Experimental results show that each algorithm presents selected trade-offs for different optimization goals, with no current approach achieving comprehensive optimization across all three metrics simultaneously.
\par
In this paper, we propose a cluster-based memory allocator, called ClusterTag, to simultaneously mitigate tag collision probability in both temporal and spatial dimensions. The fundamental insight driving ClusterTag is intuitive: current tag assignment strategies fail to optimize the aforementioned metrics (minimum, average, and unpredictability) because they cannot bridge the gap between complex memory states and limited tag quantities. Leveraging this insight, ClusterTag designs an innovative grouped memory management approach, where every 256 heap objects are concentrated in a continuous memory space called ``cluster''. This architecture confines tag assignment to the limited heap objects within clusters, making it easier to develop strategies that balance the three optimization metrics while maintaining lightweight performance. To further mitigate tag collisions between clusters, we introduce cluster-grained heap randomization. This approach cleverly improves the probabilistic protection of sanitizers through 64-bit address space, while allowing for adjustable density and effectively preventing memory fragmentation.
\par
The first goal of ClusterTag is to introduce constrained heap randomization into cluster management, seeking a balance between security and memory overhead. To this end, ClusterTag designs a three-layer memory hierarchy: \textbf{Region} (1TB) - \textbf{Pool} (1GB) - \textbf{Cluster} (256 chunks), with randomization introduced between each layer. As the fundamental randomization unit, a cluster consists of 256 contiguous chunks of uniform size, guaranteed to align with memory page boundaries. Unlike traditional heap randomization\cite{diehard,dieharder,renqing}, this design effectively prevents memory fragmentation, thereby maintaining stable memory overhead across different randomization densities. When a new cluster is allocated, ClusterTag detects whether the cluster density in the \textit{pool} exceeds the threshold. If so, it randomly allocates a new \textit{pool} within the \textit{region}. This dual randomization strategy significantly reduces page table overhead compared to distributing clusters directly across the entire \textit{region}.
\par
The second goal of ClusterTag is to design a lightweight intra-cluster tag assignment strategy, seeking to simultaneously optimize the three tag collision distance metrics within the cluster. To achieve this, ClusterTag limits each cluster to no more than 256 memory objects, thereby ensuring that 8-bit encoding can assign a spatially unique tag to each object. From the temporal dimension, ClusterTag designs a novel circular shift-based tag assignment strategy. It first randomly selects a cluster for reuse, then performs tag swapping on the freed chunks through circular right shift operations. This design not only maintains the uniqueness of tags within each cluster but also effectively extends the time window for historical tag reuse.
\par
The performance evaluation of ClusterTag was conducted on SPEC CPU 2017 \cite{cpu2017} and web servers \cite{apachebench}, with HWASan \cite{hwasan} as the baseline, which is the first widely deployed tag-based sanitizer. Results show that ClusterTag incurs similar runtime overhead and memory usage to HWASan (\textasciitilde1\%). Importantly, as randomization density decreases, ClusterTag maintains stable performance while steadily reducing tag collision probability. For security evaluation, we quantified tag collision probability across three dimensions: minimum distance, average distance, and unpredictability. The results demonstrate that ClusterTag achieves balanced optimization across all three metrics and significantly outperforms widely adopted random tag assignment strategy on most metrics. To further validate its effectiveness, we conducted a comparative analysis of vulnerability detection capabilities between ClusterTag and three existing tag assignment strategies on the Juliet \cite{juliet} dataset. Across 500 rounds of repeated testing across 7,182 vulnerability cases, ClusterTag exhibited deterministic detection/miss behavior, while other strategies presented probabilistic false negatives in 240 to 2,596 cases due to tag collisions.
\par
In a nutshell, we make the following key contributions:
\begin{itemize}
    \item We systematically analyze the limitations of tag-based sanitizers from both hardware tag encoding and software tag allocation perspectives. Attackers can exhaust their entropy through two types of tag probing attacks.
    
    \item We propose ClusterTag, a novel cluster-based memory allocator that effectively bridges the gap between chunk quantity and tag space. By grouped memory management, ClusterTag significantly mitigates tag collision probability across both spatial and temporal dimensions.
    
    \item Our evaluation sheds light on probabilistic false negatives hidden in previous tag-based sanitizers. ClusterTag can be flexibly integrated with these systems by replacing the memory allocator, substantially enhancing probabilistic protection guarantees while maintaining similar overhead.
\end{itemize}

\noindent\textbf{Availability.} To foster future research, we have released our prototype at https://github.com/Yiruma96/ClusterTag-repo.git.

\section{Background \& Related Work}
Constrained by the limited tag-encoding space, tag-based sanitizers struggle to provide sufficient probabilistic guarantees, both for in-house vulnerability detection and for in-production security protection. There are two intuitive approaches to mitigate this problem: expanding the tag-encoding space at the hardware level or optimizing tag assignment at the software level. In this section, we analyze the practical limitations of both approaches and describe how attackers can exhaust tag entropy through two types of probing attacks.

\FloatBarrier
\subsection{From ASan to Tag-based Sanitizers}
AddressSanitizer (ASan) \cite{asan} is a widely adopted tool for detecting vulnerabilities during the software testing stage. It inserts ``Red Zones'' around memory objects to check whether pointers are out of bounds (spatial error) or whether the target address is valid (temporal error). However, ASan does not associate pointers with their specific target objects. As a result, a pointer of \textit{object A} can legally point inside \textit{object B} without triggering an alert, as long as it does not encroach upon the ``Red Zone''. To overcome this limitation, tag-based sanitizers \cite{hwasan,cho2022vik,pacmem} establish a mapping relationship between pointers and their target objects through a ``key-lock'' mechanism: the ``lock'' resides in shadow memory, while the ``key'' is stored in the pointer's upper bits and automatically propagated through pointer arithmetic, ultimately enabling matching checks before memory access. 
\par
Several chip manufacturers are actively developing hardware support for tag-based sanitizers. Specifically, ARM and Intel have extended their hardware memory management units with TBI and LAM respectively, to automatically perform tag masking before pointer dereferencing. The latest ARMv9-a architecture introduces MTE, which provides dedicated tag memory isolated from normal memory, and performs safety checks in synchronous or asynchronous modes before memory access. Benefiting from these hardware features, MTE achieves low performance overhead (\textasciitilde5\% \cite{stickytag}) while eliminating the need for binary instrumentation, making it effective for protecting closed-source binary programs.

\FloatBarrier
\subsection{Hardware Tag Encoding Constraints}
\begin{figure}[h]
  \centering
  \vspace{-16.8cm}
  \setlength{\leftskip}{-9pt}
  \includegraphics[width=12in]{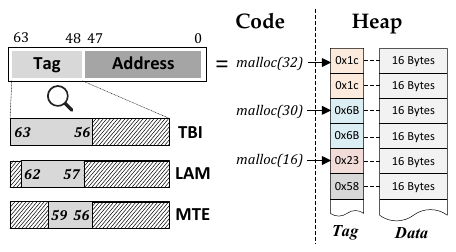}
  \vspace{-1cm}
  \caption{Tag-based Sanitizer.}
  \label{ct_hwtag}
  \vspace{-0.3cm}
\end{figure}

As shown in Figure \ref{ct_hwtag}, three mainstream hardware-level tagging mechanisms-TBI \cite{tbi_hw}, LAM \cite{lam_hw}, and MTE \cite{mte_hw}-limit the available tag space to 8 bits, 6 bits, and 4 bits, respectively. This constraint directly impacts the probabilistic protection capabilities of tag-based sanitizers. Given a tag size of $TS$ bits, the probability of successfully detecting memory vulnerabilities can be calculated as $(2^{TS} - 1)/2^{TS}$, which are 99.61\%, 98.44\%, and 93.75\% for TBI, LAM, and MTE, respectively. To reduce the probability of tag collisions, Multi-Tag \cite{multitag} extends tag encoding to the high 24 bits of pointers. However, this design relies on heavy hardware modifications, and excessive tag occupancy may also cause compatibility issues with other hardware mechanisms. xTag \cite{xtag} supports arbitrary-length pointer tags through page aliasing, but it is limited by TLB cache pressure and can use at most 4 bits (8 bits would incur a worst-case performance overhead of 251\%).

By analyzing the design documentation and community discussions of mainstream hardware mechanisms (TBI, LAM, and MTE), we attribute such conservative tag encoding to the following three constraints. These constraints reflect trade-offs made by hardware designers to ensure compatibility with existing mechanisms, although this inevitably sacrifices security to some extent.

\begin{itemize}
\item \textbf{Distinguishing kernel and user space.} The 63rd pointer bit usually separates user space (0) from kernel space (1). Using this bit for metadata is risky because the system might mistake user addresses for kernel addresses, leading to security problems. To avoid this, LAM limits memory tagging to the 62nd bit, and MTE limits it to the 59th bit.

\item \textbf{Future expansion of address space.} Support for five-level page tables continues to advance in Intel CPUs \cite{intel5} and the Linux community \cite{linux5}, which allows processes to access virtual addresses of up to 57 bits. As a result, Intel LAM limits pointer tagging to start at the 57th bit.

\item \textbf{Tagged memory overhead.} Most sanitizers store an 8-bit tag per 16 bytes of memory, resulting in 6.25\% memory overhead for tagging. Due to performance trade-offs, MTE limits tag size to 4 bits, thereby reducing memory overhead or borrowing dedicated hardware such as ECC memory \cite{ecctag1, ecctag2} for tag storage.
\end{itemize}

\FloatBarrier
\subsection{Software Tag Assignment Constraints}
\subsubsection{Definition of Tag Collision Distance}\quad \\
Given the inherent challenges in expanding hardware tag encoding, memory allocators need to carefully assign tags at the software level to reduce the possibility of collisions. Ideally, a tag assignment strategy should simultaneously optimize collision distance in both spatial and temporal dimensions for each memory tag. Spatially, the strategy must ensure that at any given moment, memory objects sharing the same tag are placed as far apart as possible. Temporally, it must ensure that for any specific memory address, the reuse of the same tag is delayed by a sufficiently long interval.
\par
To quantify the collision resistance capability of existing schemes, we model the tag assignment task as a multi-objective optimization problem. Let $\mathcal{L} = \{L_1, L_2, \ldots, L_n\}$ be the set of available tags, where $L_i$ represents the $i$-th tag. We consider the assignment of these tags to memory positions $p$ over time $t$. 
\par
For each time point $t$ and tag $L_i$, we define $P(t, L_i)$ as the ordered set of memory positions assigned the tag $L_i$ at a specific time $t$. Similarly, for each memory position $p$ and tag $L_i$, we define $\Gamma(p, L_i)$ as the ordered set of time points when memory position $p$ was assigned the tag $L_i$. Under these definitions, the multisets of adjacent spatial collision distances $\mathcal{D}_S$ and temporal collision distances $\mathcal{D}_T$ can be represented by Equation (\ref{eq:ds}) and Equation (\ref{eq:dt}). Here, $p_j$ and $t_j$ represent the $j$-th elements in the ordered sets $P(t, L_i)$ and $\Gamma(p, L_i)$, respectively. The symbol $\uplus$ denotes the multiset sum, which combines all collections of distances while preserving duplicate values.

\begin{equation}
\mathcal{D}_S = \biguplus_{t, i} \{ |p_{j+1} - p_j| : j=1, \ldots, |P(t, L_i)|-1 \} \label{eq:ds}
\end{equation}

\begin{equation}
\mathcal{D}_T = \biguplus_{p, i} \{ |t_{j+1} - t_j| : j=1, \ldots, |\Gamma(p, L_i)|-1 \} \label{eq:dt}
\end{equation}

Following the aforementioned definitions, we establish three optimization objectives for each type of tag assignment strategy:
\begin{itemize}
  \item \textbf{Objective 1: Increase the minimum collision distances.} The strategy should prevent assigning the same tag to adjacent memory objects and avoid reassigning a recently released tag to the same memory address. This can be formalized as maximizing $f_1 = \min(\mathcal{D})$, where $\mathcal{D}$ represents either $\mathcal{D}_S$ or $\mathcal{D}_T$.
  
  \item \textbf{Objective 2: Increase the average collision distances.} The strategy should ensure substantial spatial and temporal separation between memory objects sharing the same tag, thereby reducing the impact range of out-of-bounds accesses and dangling pointer errors. This corresponds to maximizing $f_{2} = \text{avg}(\mathcal{D})$.
  
  \item \textbf{Objective 3: Increase the unpredictability of collision distances.} 
  The statistical distributions of $\mathcal{D}_S$ and $\mathcal{D}_T$ should provide sufficient entropy to prevent attackers from predicting which memory objects share the same tag. We quantify this by maximizing $f_{3} = H(\mathcal{D})$, where $H()$ is the entropy function.
\end{itemize}

\FloatBarrier
\subsubsection{Existing Tag Assignment Strategies in Allocators} \quad\\
Although several approximation algorithms \cite{multiobjective1, multiobjective2} exist to solve the above multi-objective optimization problem, they are too heavy for memory allocators and thus difficult to apply in real-time scenarios. 1) First, calculating temporal collision distances requires allocators to record historical tag sequences for each memory address, resulting in additional memory overhead. 2) Second, modern applications perform tens of thousands of memory allocations per second, and calculating collision distance in real time would significantly degrade memory throughput and overall performance. 3) Finally, the algorithm cannot predict future memory allocations, which means that optimal tag assignments for current objects may negatively impact subsequent objects, preventing globally optimal solutions.
\par
Due to these practical limitations, existing memory allocators typically adopt lightweight tag assignment strategies to accommodate high memory throughput. Table \ref{allocator} summarizes three representative tag assignment strategies-random, interleaved, and fixed-employed by five mainstream allocators. We abstracted the tag distributions from their specific algorithmic implementations and calculated their values based on the three optimization functions defined above. Our analysis reveals that each strategy emphasizes different optimization objectives, making them suitable for defending against distinct vulnerability types.

\begin{table}[h]
  \centering
  \caption{Comparison of three types of tag assignment strategies (TS=Tag Size: HWASan=8, Others=4)}
  \label{allocator}

  \small
  \setlength{\tabcolsep}{2.35pt}
  \renewcommand{\arraystretch}{0.95}  

  \begin{tabular}{l  l | cc  cc  cc}
      \toprule
      \multirow{3}{*}{\textbf{Strategy}} & \multirow{3}{*}{\textbf{Research}} & \multicolumn{2}{c}{\textbf{Objective1}} & \multicolumn{2}{c}{\textbf{Objective2}} & \multicolumn{2}{c}{\textbf{Objective3}} \\
      & & \multicolumn{2}{c}{(Minimum)} & \multicolumn{2}{c}{(Average)} & \multicolumn{2}{c}{(Unpred.)} \\
      \cmidrule(lr){3-4} \cmidrule(lr){5-6} \cmidrule(lr){7-8}
      & & \textit{Space} & \textit{Time} & \textit{Space} & \textit{Time} & \textit{Space} & \textit{Time} \\

      \midrule

      \noalign{\vskip 3pt} 
      \multirow{2}{*}{\textbf{Random}} & HWASan \cite{hwasan}            & \textit{1}        & \textit{1}        & \textit{$2^{TS}$} & \textit{$2^{TS}$}   & \textit{9.44} & \textit{9.44} \\[-2pt]
      \noalign{\vskip 3pt} 
                                       & Ptmalloc \cite{ptmalloc}        & \textit{header}   & \textit{1}        & \textit{$2^{TS}$} & \textit{$2^{TS}$}   & \textit{5.40} & \textit{5.40} \\[-2pt]
      \noalign{\vskip 3pt} 
      \midrule
      \noalign{\vskip 3pt} 
      \textbf{Staggered}               & Scudo \cite{scudo}              & \textit{2}        & \textit{2}        & \textit{$2^{TS}$} & \textit{$2^{TS-1}$} & \textit{4.35} & \textit{4.35} \\[-2pt]
      \noalign{\vskip 3pt}
      \midrule
      \noalign{\vskip 3pt}
      \multirow{2}{*}{\textbf{Fixed}}  & PartitionAlloc \cite{partition} & \textit{1}        & \textit{$2^{TS}$} & \textit{$2^{TS}$} & \textit{$2^{TS}$}   & \textit{5.40} & \textit{0}    \\[-2pt]
      \noalign{\vskip 3pt}
                                       & StickyTags \cite{stickytag}      & \textit{$2^{TS}$} & \textit{1}        & \textit{$2^{TS}$} & \textit{1}          & \textit{0}    & \textit{0}    \\[-2pt]
      \noalign{\vskip 3pt}
      
      \bottomrule
  \end{tabular}
\end{table}

\vspace{0.1cm}
\noindent\textbf{Random Assignment.} Memory allocators like HWASan \cite{hwasan} and Ptmalloc \cite{ptmalloc} assign random tags to memory objects, following a geometric probability distribution. While random allocation effectively increases the unpredictability of address-to-tag mapping (\textcolor{mygreen}{Objective 3}), it sacrifices the critical property of minimum collision distance (\textcolor{myred}{Objective 1}). First, spatially adjacent memory objects may be randomly assigned identical tags, allowing attackers to exploit adjacent overflow vulnerabilities. Second, memory addresses may be assigned the same tag as in the previous round, making temporal errors easier to exploit. It should be noted that Ptmalloc embeds the \textit{header} at the beginning of each chunk and assigns it a fixed tag of 0. This memory layout allows Ptmalloc to deterministically detect adjacent overflow vulnerabilities.

\vspace{0.1cm}
\noindent\textbf{Staggered Assignment.} Scudo\cite{scudo} divides its available tag set into two distinct groups: one for odd-indexed allocations and one for even-indexed allocations. The tag within each allocation is randomly selected, thus following a geometric probability distribution. While this staggered allocation strategy enables deterministic detection of adjacent overflow vulnerabilities (\textcolor{mygreen}{Objective 1}), it comes with notable trade-offs. First, it reduces the average collision distance (\textcolor{myred}{Objective 2}) in the temporal dimension to $2^{TS-1}$. Second, reducing the tag space by half also reduces the entropy of collision distances (\textcolor{myred}{Objective 3}). As shown in Table \ref{allocator}, under the identical 4-bit tag size, Scudo provides lower entropy than Ptmalloc.
\par
Vulnerability tracking by the Microsoft Security Team has shown that the proportion of non-adjacent overflow vulnerabilities has increased in recent years \cite{microvultypes}. However, the three memory allocators Scudo, HWASan, and Ptmalloc randomly assign tags to memory objects, causing their collision distances to follow a geometric distribution. This distribution means that most collision distances are short, which significantly reduces their effectiveness in protecting against non-adjacent overflows.

\vspace{0.1cm}
\noindent\textbf{Fixed Assignment.} This type of strategy is dedicated to improving the minimum collision distance (\textcolor{mygreen}{Objective 1}) to effectively detect memory vulnerabilities such as non-adjacent overflows, but it typically makes compromises on predictability (\textcolor{myred}{Objective 3}). Specifically, PartitionAlloc \cite{partition} increments the tag when an object is released, thereby increasing the temporal collision distance to $2^{TS}$. However, when attackers can manipulate the timing of allocation and deallocation, they can successfully construct illegal pointers with the same tag as the target object. In contrast to PartitionAlloc, StickyTags \cite{stickytag} focuses on the minimum collision distance in the spatial dimension. This work assigns incremental persistent tags to objects, resulting in zero entropy for collision distance in both spatial and temporal dimensions. Therefore, attackers can not only accurately predict memory addresses with the same tag, but can also arbitrarily exploit temporal vulnerabilities.

\FloatBarrier
\subsection{Two Types of Tag Probing Attacks}
This subsection introduces two types of tag probing approaches that enable attackers to perform brute force attacks without disrupting the target system's usability. Through these approaches, an attacker can gradually exhaust the entropy of the tag space and eventually guess the correct (but illegal) tag pair.

\vspace{0.1cm}
\noindent\textbf{Crash recovery-based tag probing scheme.} Web servers \cite{nginx} and browsers \cite{firefox2008mozilla} adopt a Master-Worker architecture to achieve high availability and memory isolation. In this architecture, the Master process serves as a checkpoint of the initial state and is responsible for restarting Worker processes through \textit{fork} when they unexpectedly crash. Previous research \cite{bittau2014hacking,gawlik2016enabling} has demonstrated that attackers can maliciously exploit this crash recovery feature to exhaust the entropy of probabilistic protection schemes (such as ASLR). Exploiting this same recovery mechanism, attackers can repeatedly probe memory objects within Worker processes and gradually leak the tag by crash/no-crash feedback.

\vspace{0.1cm}
\noindent\textbf{Speculative execution-based tag probing scheme.} Even with non-crash resistant programs, attackers can probe memory tags covertly through side-channel attacks. StickyTags \cite{stickytag} and TikTag \cite{kim2024tiktag} demonstrated how speculative execution vulnerabilities \cite{kocher2020spectre} impact real MTE hardware. In synchronous mode, MTE immediately blocks all illegal memory accesses and results in a 0\% cache hit rate, allowing attackers to infer tag matches indirectly through cache hit status. In asynchronous mode, although MTE delays program termination to obscure tag matching results, researchers observed that the accumulation of illegal memory accesses gradually shortens this delay window. This phenomenon means that the probability of cache hits progressively decreases after multiple failed matches, enabling attackers to infer tag matching results even in MTE's asynchronous mode.

\section{Threat Model}
Throughout this paper, we use ClusterTag as an in-house sanitizer for mining heap-related memory vulnerabilities. The specific threat model is as follows:

\vspace{0.1cm}
\noindent\textbf{Vulnerability Assumptions.} We assume the existence of heap memory vulnerabilities in programs, including spatial (off-by-N, type confusion) and temporal (double free, use-after-free) vulnerabilities. These memory violations are typically concentrated near illegal pointers; hence, increasing the tag collision distance can enhance the effectiveness of vulnerability detection. We consider other memory violations, such as memory leaks, uninitialized memory, and undefined behavior, to be out of scope, as they are the subject of extensive literature on orthogonal defenses \cite{msan, lsan, ubsan}.

\vspace{0.1cm}
\noindent\textbf{Attacker Assumptions.} Considering the potential of integrating ClusterTag with MTE, we also evaluate the attack surface of ClusterTag when used for in-production exploit mitigation. We assume that attackers can perform a limited number of attempts through the two aforementioned tag probing attacks and can implement heap spraying through memory deallocation/allocation operations. We assume that attackers cannot directly manipulate tags, as most vulnerability types, such as temporal vulnerabilities (use-after-free, double free) and constrained out-of-bounds accesses (off-by-N, type confusion), do not provide this capability. Regarding randomization, we assume that attackers understand the functionality of ClusterTag but do not know the randomization seed.

\section{Design}
This section introduces ClusterTag, a cluster-based memory allocator designed to effectively mitigate tag collisions between memory objects. We first outline the workflow, starting from inter-cluster memory management and extending to intra-cluster tag assignment. Subsequently, we detail the three key components of ClusterTag, discussing their respective design challenges and proposed solutions.

\begin{figure}[t]
  \centering
  \vspace{-13.4cm}
  \setlength{\leftskip}{-19pt}
  \includegraphics[width=11.6in]{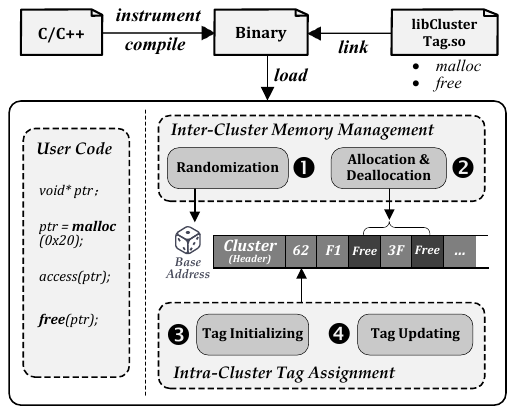}
  \vspace{-1cm}
  \caption{Overview of ClusterTag.}
\vspace{-0.2cm}
  \label{ct_overview}
\end{figure}

\subsection{Overview}
As shown in Figure \ref{ct_overview}, ClusterTag functions as an independent memory allocator linked to the target program, providing standard memory management interfaces (e.g., \textit{malloc}, \textit{free}). This design enables ClusterTag to maintain backward compatibility for closed-source programs without requiring additional instrumentation. Furthermore, it enables flexible combination with various tag-based sanitizers \cite{hwasan} and provides inherent tag collision mitigation capabilities for them. Following the cluster-based partitioning principle, ClusterTag's overall architecture comprises two primary components, as shown in Figure \ref{ct_overview}: one responsible for inter-cluster memory management and the other for intra-cluster tag assignment. These components will be discussed separately below.

\vspace{0.1cm}
\noindent\textbf{Inter-Cluster Memory Management.} ClusterTag introduces cluster-grained randomization to mitigate tag collisions across clusters (\ding{182}). Specifically, ClusterTag employs a carefully designed three-layer memory layout, referred to as the \textit{Region-Pool-Cluster}, with randomization applied between layers. This hierarchical structure avoids the drawbacks that randomization typically imposes on memory locality, thus enhancing memory page utilization and reducing page table overhead, particularly in scenarios with low-density randomization. For efficient cluster-based runtime memory management (\ding{183}), ClusterTag utilizes doubly-linked lists to organize all clusters, enabling reuse of freed chunks to minimize fragmentation, and periodically returning contiguous free pages to the kernel.

\vspace{0.1cm}
\noindent\textbf{Intra-Cluster Tag Assignment.} Benefiting from the mitigation of cross-cluster tag collisions through randomization, the tag assignment task (\ding{184}, \ding{185}) can focus on limited chunks within a single cluster. Specifically, ClusterTag assigns unique tags to each chunk within an initialized cluster, which can inherently eliminate intra-cluster tag collisions in the spatial dimension. When clusters are reused, ClusterTag designs a circular-shift tag updating strategy to optimize tag collisions in the temporal dimension. Apart from improving security, ClusterTag's tag assignment strategy is extremely lightweight and does not require any additional space to record historical tags.

\FloatBarrier
\subsection{Cluster-grained Heap Randomization}
\begin{figure}[t]
    \centering
    \vspace{-13.7cm}
    \setlength{\leftskip}{-8pt}
    \includegraphics[width=12in]{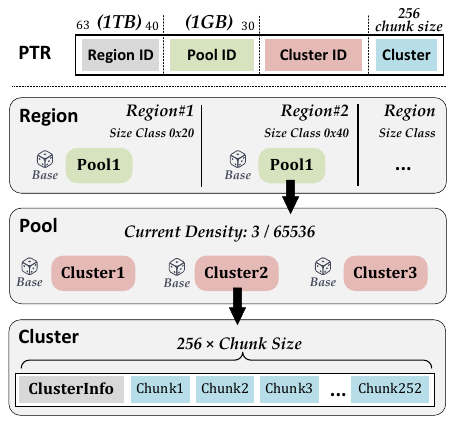}
    \vspace{-0.8cm}
    \caption{Three-layer Heap Randomization.}
    \vspace{-0.4cm}
    \label{clusterrandomization}
\end{figure}

ClusterTag employs a cluster-grained heap randomization scheme to alleviate tag collisions between memory objects located in different clusters. While traditional heap randomization techniques \cite{diehard,dieharder} are often used to mitigate memory vulnerability exploitation, they typically break memory locality and introduce significant memory overhead. On the one hand, sparse data units fail to fully utilize individual physical pages, resulting in memory fragmentation. On the other hand, the random distribution of data units also requires more page table entries (PTEs) for address translation. To overcome these challenges, ClusterTag first adopts a BiBOP (Big Bag of Pages) \cite{hwasan,diehard,duck2016heap,freebsd} style memory organization strategy. It allocates same-sized chunks contiguously within a dedicated \textit{region}, ensuring proper alignment and contiguous memory page occupation from a localized perspective. Unlike typical BiBOP-style allocators, ClusterTag innovatively introduces cluster-grained randomization within each region, creating randomized gaps between chunks from a global perspective. As illustrated in Figure \ref{clusterrandomization}, ClusterTag constructs a three-layer memory layout (\textit{Region-Pool-Cluster}) and introduces variable-density randomization across these layers. Design details for each layer are described in the following parts:

\vspace{0.1cm}
\noindent\textbf{Cluster.} A cluster consists of a group of contiguous chunks with identical size, serving as the basic unit for both heap randomization and tag assignment. Selecting an appropriate number of memory chunks per cluster requires balancing two considerations: matching the available tag space (avoiding too many chunks) and utilizing complete physical memory pages (avoiding too few chunks). To achieve this balance, ClusterTag restricts the number of chunks per cluster to 256, ensuring each chunk can be assigned a unique tag within an 8-bit tag space. Furthermore, ClusterTag defines 30 size classes across 30 TB of virtual space, with sizes growing in piecewise-linear steps: \textit{0x20}, \textit{0x40}, ..., \textit{0x100}, \textit{0x200}, \textit{0x300}, ..., up to \textit{0x10000}. Each size class is an integer multiple of \textit{0x10}; therefore, the cluster size is standardized to $256 \times (0x10 \times N) = 4096 \times N$ bytes. This design guarantees that clusters can be strictly aligned with physical memory pages, effectively avoiding memory waste issues caused by heap randomization. 
\par
To prevent object metadata corruption by adjacent overflows, we store all metadata centrally in the \textit{ClusterInfo} structure at each cluster's head, separated from memory objects. The \textit{ClusterInfo} tracks each object's status: 0 for in-use objects, non-zero for free objects (storing the previous assignment's tag). We assign \textit{ClusterInfo} a fixed tag of 0 to prevent malicious tampering by tagged pointers.

\vspace{0.1cm}
\noindent\textbf{Pool.} In modern 64-bit architectures, a single page can store 512 page table entries, enabling it to address 2MB of contiguous memory space. However, randomization typically cannot effectively utilize this contiguous space, resulting in numerous idle page table entries. To optimize page table overhead, ClusterTag introduces the concept of \textit{pool}, which constrains the randomization range of clusters within a \textit{region}. Each \textit{pool} has a fixed size of 1GB, allowing the kernel to complete addressing using a single level-3 page table entry and 512 level-4 page table entries (\textasciitilde2MB). This level of memory overhead is acceptable for most modern devices. 
\par
Additionally, ClusterTag defines a configurable randomization density parameter $d$ to limit the maximum cluster quantity in each \textit{pool}. The primary experiments in our paper were conducted with $d=5$, which means only $1\text{GB}/5 = 205$ MB of virtual addresses is used in each 1GB \textit{pool}. Once a \textit{pool} reaches its density limit, ClusterTag dynamically allocates a new \textit{pool} at a randomized location within the designated \textit{region}. This dual randomization strategy not only ensures randomness in cluster distribution but also significantly optimizes page table overhead, achieving a balance between security and memory efficiency.

\vspace{0.1cm}
\noindent\textbf{Region.} ClusterTag allocates all chunks of identical size within a contiguous 1TB memory space, referred to as a \textit{region}. This BiBOP-style memory organization offers two key advantages. First, for any pointer $PTR$, ClusterTag can quickly determine its corresponding \textit{region} id by computing $(PTR >> 40) \& 0xFF$ and further identify the size class of that chunk, which facilitates efficient deallocation operations. Second, this memory organization separates headers from chunks, allowing ClusterTag to centralize them within the \textit{ClusterInfo}, as shown in Figure \ref{clusterrandomization}. Our design intentionally avoids allocating all available slots in the cluster for memory objects, thereby reserving certain tags in the 8-bit encoding space. These reserved tags serve as a buffer during intra-cluster tag assignment to extend the minimum circular chain length. We will describe this design in detail in subsection 4.4.

\FloatBarrier
\subsection{Cluster-based Memory Alloc \& Dealloc}
Based on the new cluster-based memory model, ClusterTag further designs a runtime memory allocation and deallocation scheme. Similar to other memory allocators, the primary goal of this scheme is to reduce memory fragmentation to minimize physical memory usage. Secondly, the scheme aims to shorten the average lifetime of clusters as much as possible, thereby reducing the probability of temporal tag collisions. To achieve these two objectives, ClusterTag organizes the clusters within each \textit{Region} as a doubly linked list, as shown in Figure \ref{clustermanagement}, and the detailed runtime memory allocation and deallocation scheme is described next.

\subsubsection{Memory Allocation Scheme}
\begin{figure}[t]
  \centering
  \vspace{-18.5cm}
  \setlength{\leftskip}{-15pt}
  \includegraphics[width=14.7in]{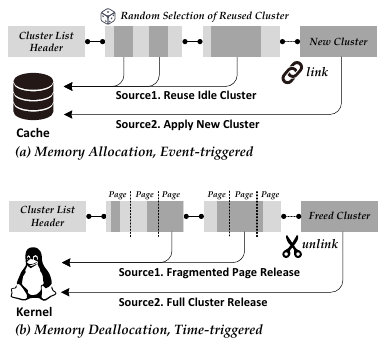}
  \vspace{-1.3cm}
  \caption{Runtime Memory Allocation and Deallocation.}
\vspace{-0.1cm}
  \label{clustermanagement}
\end{figure}

\quad \\
Similar to other BiBOP-style memory allocators \cite{dieharder,hwasan,freeguard}, ClusterTag directly allocates large objects via the \textit{mmap} system call to reduce memory fragmentation, with their base addresses randomized by the kernel's ASLR. The threshold for large objects is empirically set at 64KB (0x10000 bytes), consistent with our baseline HWASan \cite{hwasan} and other work \cite{freeguard}. For memory objects smaller than 64KB, ClusterTag first aligns the requested size and then returns a pre-allocated freed heap chunk from the cache array. As illustrated in Figure \ref{clustermanagement}, when the cache array becomes exhausted, ClusterTag will refill it from the following two sources.

\vspace{0.1cm}
\noindent\textbf{Reusing idle clusters.} ClusterTag preferentially fills cache arrays with idle heap clusters to reduce runtime intra-cluster memory fragmentation. Unlike traditional memory allocators that tend to reuse recently freed heap chunks, ClusterTag randomly selects from all idle clusters and reuses their internal freed chunks, which further enhances the unpredictability of their internal states on top of cluster base address randomization. Additionally, ClusterTag treats clusters rather than chunks as the basic reuse unit. This approach concentrates cached chunks within specific clusters instead of dispersing them across many different clusters, which allows unselected clusters adequate time to be fully released. Subsequent experiments demonstrate that this design significantly shortens the lifetime of clusters, allowing most of them to be returned to the kernel before internal temporal tag collisions occur.

\vspace{0.1cm}
\noindent\textbf{Applying new cluster.} When the number of freed chunks in all clusters is insufficient to fill the cache array, ClusterTag requests a new cluster from the kernel. Specifically, ClusterTag first selects a \textit{pool} within the \textit{region} that has not reached the density threshold, then randomly partitions a contiguous memory space of twice the cluster size within that \textit{pool}. ClusterTag assigns only the first half of the requested space to the new cluster. This strategy ensures that adjacent clusters always maintain a minimum distance of $256 \times chunk\_size$, thereby increasing the minimum spatial distance for cross-cluster tag collisions.

\subsubsection{Memory Deallocation Scheme}\quad \\
When a user process releases a memory block through \textit{free}, ClusterTag also first determines whether the requested size exceeds 64KB. For large objects, ClusterTag directly releases them through the \textit{munmap} system call. For memory objects smaller than 64KB, ClusterTag marks them as freed within the cluster header (\textit{ClusterInfo}). These freed heap chunks are either reused when filling the cache arrays or returned to the kernel through periodic cluster chain scans.
\par
As shown in Figure \ref{clustermanagement}, periodic scanning includes two types: full release and fragmented release. When all memory blocks in a cluster have been freed, ClusterTag returns memory pages occupied by the cluster to the kernel and removes it from the cluster chain. This design effectively shortens the lifetimes of clusters, making the distribution of memory objects more random. If some chunks within the cluster remain in use, ClusterTag will further check whether the freed chunks form contiguous pages and return such space to the kernel. The page threshold that triggers fragmented release is configurable to avoid performance overhead caused by excessive interactions between ClusterTag and the kernel.

\FloatBarrier
\subsection{Intra-Cluster Tag Assignment}
\begin{figure}[b]
  \centering
  \vspace{-5.7cm}
  \setlength{\leftskip}{8pt}
  \includegraphics[width=7in]{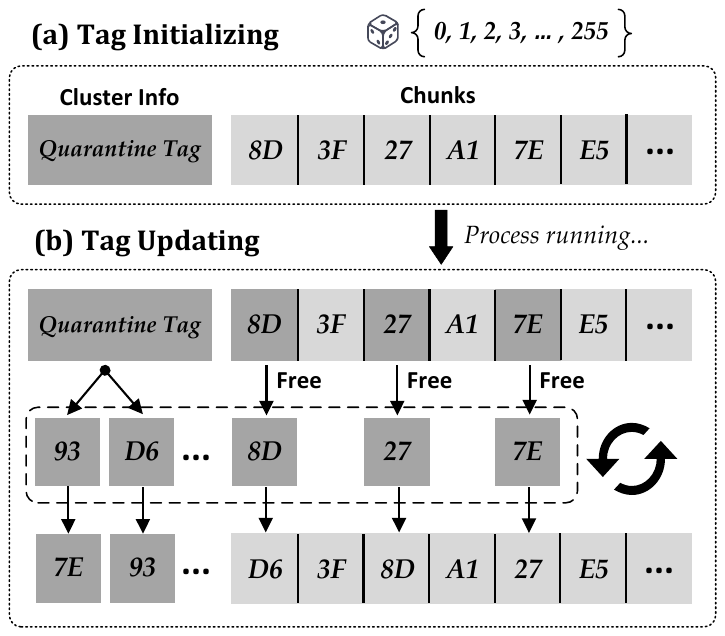}
  \vspace{-0.4cm}
  \caption{Circular-shift Tag Assignment Strategy.}
\label{tagassignment}
\end{figure}

Building upon the cluster-based memory management, ClusterTag effectively simplifies the tag assignment task to a small number of intra-cluster chunks. Besides, it also eliminates the uncertainty of the memory allocation order within the cluster, thereby enabling ClusterTag to globally optimize tag assignment for all allocated chunks. Building on these advantages, ClusterTag carefully designs the tag assignment scheme shown in Figure \ref{tagassignment}, which is used both for tag initializing of new clusters and tag updating of reused clusters.

\vspace{0.1cm}
\noindent\textbf{Tag initializing for new cluster.} The current design of ClusterTag employs 8-bit tag encoding, ensuring that all chunks within a cluster can be assigned unique tags upon the first tag assignment, which completely eliminates spatial tag collisions. Furthermore, ClusterTag reserves certain free tags without assigning them to any objects, as illustrated in Figure \ref{tagassignment}. These quarantine tags are derived from memory space occupied by the \textit{ClusterInfo} structure, and they will be used to improve the subsequent tag updating task.

\vspace{0.1cm}
\noindent\textbf{Tag updating for reused cluster.} ClusterTag designs a lightweight tag reassignment strategy for freed heap chunks based on the idea of circular shift. As shown in Figure \ref{tagassignment}, ClusterTag organizes quarantine tags (\textit{93}, \textit{D6}) along with tags from previously freed chunks (\textit{8D}, \textit{27}, \textit{7E}) into a ring, and then swaps them via a one-position circular right shift. This tag updating strategy first maintains the spatial uniqueness of tags, ensuring that no tag collisions occur within reused clusters at any time. In the time dimension, all free tags will rotate systematically within the cluster before reuse, thereby increasing the reassignment interval to the same chunk; on the other hand, the existence of quarantine tags guarantees a minimum cycle length, effectively preventing free tags from being immediately reused. In Section 5.4, we will present and compare the tag collision distance distributions for ClusterTag and HWASan (which uses a random tag assignment strategy). Furthermore, we will provide a detailed discussion of their differences regarding three key metrics: minimum distance, average distance, and unpredictability.

\FloatBarrier
\section{Evaluation}
The prototype of ClusterTag is implemented as an independent memory allocator with an 8-bit tag length, which can provide transparent temporal and spatial tag collision mitigation for upper-layer sanitizers. After introducing the experimental setup in subsection 5.1, we will conduct a detailed evaluation of ClusterTag and mainly answer the following three research questions in the subsequent subsections:

\begin{itemize}
    \item \textbf{RQ1. (Performance)} What runtime and memory overhead does ClusterTag incur, and how do these overheads evolve as the randomization density varies?
    
    \item \textbf{RQ2. (Security)} Does ClusterTag reduce tag collision probability during vulnerability detection, and can it effectively defeat active tag probing attacks?
    
    \item \textbf{RQ3. (Collision Distance Quantification)} Compared to existing tag assignment strategies, does ClusterTag achieve a better balance among the three optimization objectives for tag collision distance?
\end{itemize}

\FloatBarrier
\subsection{Experimental Setup}
\noindent\textbf{Baseline Work.} Our experiments select the Hardware-assisted Sanitizer (HWASan) in LLVM-15 \cite{hw_llvm15} as the evaluation baseline for ClusterTag. As the first widely deployed tagged memory scheme, HWASan utilizes the high 8 bits of pointers to tag memory objects and relies on ARM TBI hardware to provide pointer masking support. HWASan's implementation consists of a set of compile-time instrumentation passes and a runtime memory allocator. For consistent performance evaluation, we only replaced HWASan's native memory allocator with ClusterTag while keeping its instrumentation strategy unchanged. For security evaluation, since ClusterTag's tag collision mitigation currently only applies to heap variables, we disabled HWASan's security checks for stack variables to focus on their probabilistic protection capabilities for heap variables. We will introduce feasible extensions of ClusterTag for stack memory protection in the discussion section.

\vspace{0.1cm}
\noindent\textbf{Experimental Platform.} Since the baseline implementation (HWASan) relies on TBI hardware support, our experiments were primarily conducted on the ARM64 architecture. It should be noted that the prototype of ClusterTag is architecture-independent. With the progressive maturity of hardware-supported memory tagging technologies such as Intel LAM\cite{lam_hw}, ClusterTag can be applied to sanitizers on the x64 architecture. The experiments were conducted on an ARM64 server equipped with a Kunpeng 920 processor \cite{xia2021kunpeng} (64 CPU cores at 2.6GHz) and 512GB RAM, running on Ubuntu 22.04. To ensure the accuracy of experimental results, 32 CPU cores were isolated from the kernel's process scheduling system and dedicated exclusively for performance testing.

\FloatBarrier
\subsection{Performance Evaluation}
To answer research question \textbf{RQ1}, we evaluate ClusterTag's performance overhead using SPEC CPU 2017 benchmarks. As shown in Tables \ref{cluster2017time} and \ref{cluster2017memory}, the suite's memory usage exceeds 4GB with allocation frequencies up to 28,981 per second, making it suitable for evaluating ClusterTag's stability and resource efficiency under sustained workloads. Our evaluation includes all 16 applications from SPEC CPU 2017 written in C and C++. Four programs (627.cam4\_s, 602.gcc\_s, 620.omnetpp\_s, and 625.x264\_s) encountered runtime errors or were terminated by sanitizers due to memory exceptions, and thus were excluded from the final dataset.

\FloatBarrier
\subsubsection{SPEC CPU 2017 Runtime Overhead}\quad \\
Table \ref{cluster2017time} presents a comparison of runtime overhead between the original programs (linked with Glibc) and the hardened programs employing either HWASan or ClusterTag. The results indicate that both HWASan and ClusterTag incur approximately 80\% runtime overhead relative to the original programs \footnote{$(6,190-3,335)/3,335 \approx 85\%, \quad (6,205-3,335)/3,335 \approx 86\%$}. This additional overhead primarily stems from two factors: differences in memory management strategies and extra instrumentation instructions, including shadow memory accesses and tag matching operations. In our experiments, ClusterTag replaced only the memory allocator component of HWASan, while preserving identical compile-time instrumentation, ensuring that both approaches incur identical instrumentation-related overhead.
\par
Although both HWASan and ClusterTag employ the same compile-time instrumentation, there are significant differences in their memory management strategies. HWASan stores memory objects contiguously and assigns random tags to each object, whereas ClusterTag organizes memory objects into distinct clusters and assigns tags in a circular pattern within each cluster. As shown in Table \ref{cluster2017time}, ClusterTag's runtime overhead is slightly lower than HWASan's by approximately 1\%, which demonstrates the lightweight nature of its inter-cluster randomization and intra-cluster tag assignment schemes. 

\begin{table}[h]
    \centering
    \caption{Runtime overhead of ClusterTag on SPEC CPU 2017 ($d$=randomization density; asterisks indicate significant difference compared to HWASan: *p<0.05, **p<0.01, ***p<0.001)}  
    \label{cluster2017time}

    \renewcommand{\arraystretch}{1.15}
    \small
    \setlength{\tabcolsep}{2.7pt}
    
    \begin{tabular}{l r r | r r r | r}
\toprule
    \multirow{2}{*}{\textbf{Benchmark}} & \textbf{Glibc} & \textbf{HW} & \multicolumn{3}{c|}{\textbf{ClusterTag}} & \textbf{Alloc} \\
    \cmidrule(lr){4-6}
     & (sec) & (sec) & $(d=5)$ & $(d=10)$ & $(d=20)$ & (/sec) \\
    \midrule
    600.perlbench\_s   &   714 &  3,163 &  3,222**  &  3,218 &  3,230 &  51,367 \\
    605.mcf\_s         & 1,452 &  3,084 &  3,046    &  3,085 &  3,023 &     326 \\
    607.cactuBSSN\_s   & 3,532 &  7,852 &  8,000    &  7,958 &  8,064 &      32 \\
    619.lbm\_s         & 2,103 &  3,553 &  3,591    &  3,626 &  3,565 &       2 \\
    621.wrf\_s         & 5,064 &  6,633 &  6,571**  &  6,430 &  6,523 & 111,534 \\
    623.xalancbmk\_s   &   633 &  1,078 &  1,133*** &  1,136 &  1,106 & 128,353 \\
    628.pop2\_s        & 5,124 &  6,370 &  6,315*** &  6,249 &  6,319 &  22,565 \\
    631.deepsjeng\_s   &   478 &  1,641 &  1,641    &  1,638 &  1,637 &       1 \\
    638.imagick\_s     & 8,995 & 20,063 & 20,035    & 20,043 & 20,038 &   2,140 \\
    641.leela\_s       &   548 &  1,630 &  1,605**  &  1,601 &  1,606 &  31,314 \\
    644.nab\_s         & 6,647 & 12,070 & 12,144    & 12,117 & 12,140 &     138 \\
    657.xz\_s          & 4,727 &  7,146 &  7,160    &  7,162 &  7,086 &       1 \\
    \midrule
\textit{Average}   & 3,335 &  6,190 & \cellcolor[gray]{0.9}6,205 & \cellcolor[gray]{0.9}6,189 & \cellcolor[gray]{0.9}6,195 &  28,981 \\
    \bottomrule
    \end{tabular}
\end{table}

To determine whether the performance differences between ClusterTag and HWASan are statistically significant, we ran each benchmark five times and performed two-sample t-tests ($\alpha=0.05$). The result, summarized in Table \ref{cluster2017time} shows that 5 out of 12 test programs exhibit statistically significant differences (p < 0.05). Among these, \textit{623.xalancbmk\_s} and \textit{628.pop2\_s} show particularly strong significance (p < 0.001). This is because all five programs have memory allocation frequencies exceeding 20,000 per second, where even minor performance differences can become statistically significant. In contrast, the remaining seven programs involve relatively infrequent dynamic memory allocations, and their performance is primarily determined by compute-intensive operations. As a result, the impact of allocator-related overhead is negligible, and no statistically significant differences are observed.
\par
To further evaluate the performance impact of ClusterTag's randomization density, we conducted experiments with $d$ values of 5, 10, and 20. Higher $d$ values correspond to sparser distribution of clusters within the \textit{pool}. As shown in Table \ref{cluster2017time}, ClusterTag maintained stable performance across these configurations, with average overheads of 0.24\%, -0.02\%, and 0.08\%, respectively. This result is readily explained by the fact that each randomization unit in ClusterTag occupies a complete physical page, thus not affecting memory locality (such as reducing Cache and TLB hit rates). By contrast, traditional heap randomization research such as DieHarder \cite{dieharder} already incurs a 20\% performance overhead even when operating at a higher density (utilizing 7/8 of each page).

\FloatBarrier
\subsubsection{Detailed Performance Overhead Analysis}
\begin{figure}[h]
  \centering
\setlength{\leftskip}{-3pt}
  \includegraphics[width=3.45in]{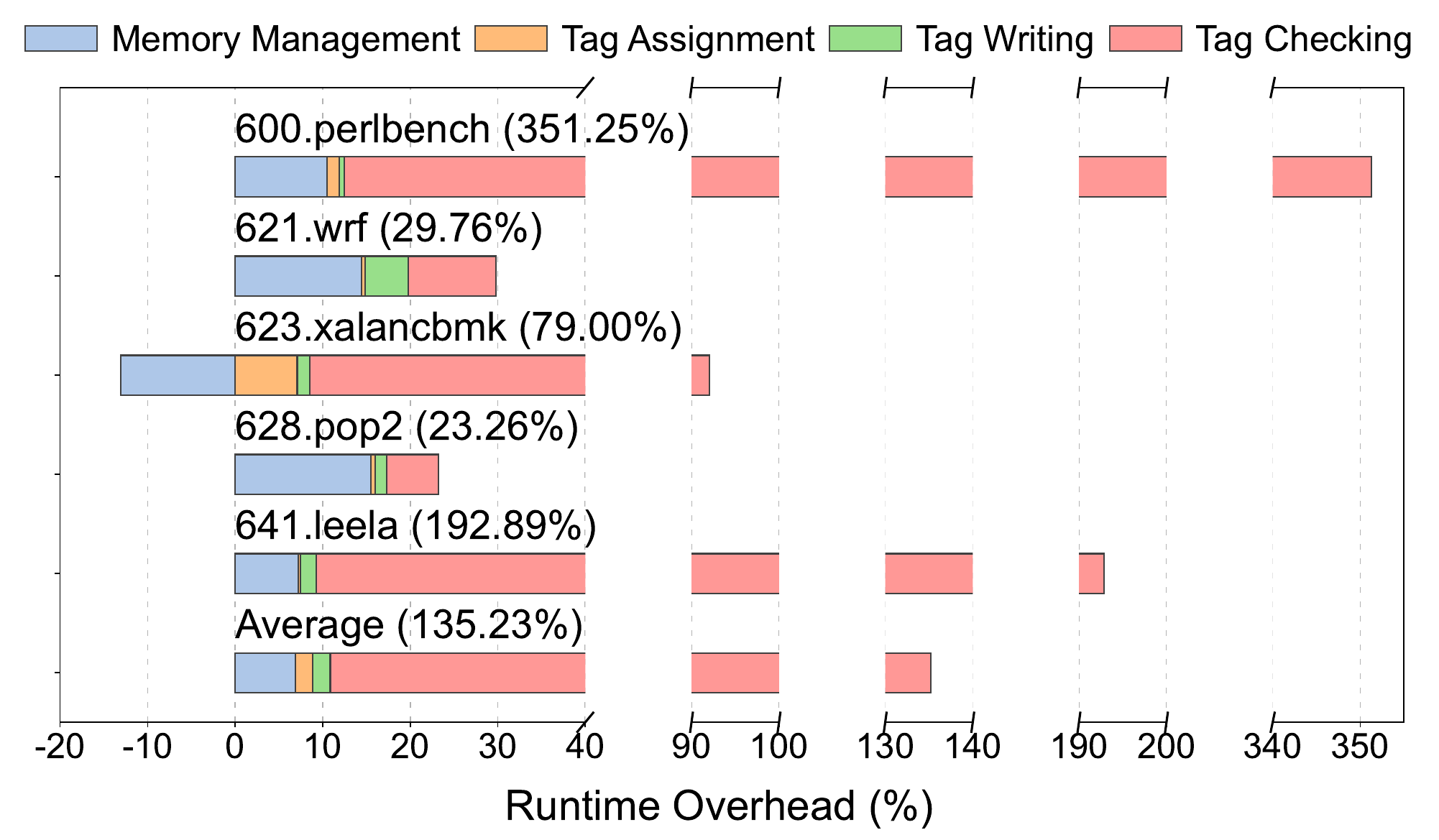}
\caption{Runtime overhead buildup on SPEC CPU 2017.}
\label{ct_buildup}
\end{figure}

\quad \\
To better understand ClusterTag's performance overhead, we conducted a detailed breakdown across four key components: memory management (Sections 4.2 and 4.3), tag allocation (Section 4.4), tag writing, and tag checking. The evaluation is based on SPEC CPU 2017 benchmarks, using Glibc-based programs as the baseline. To ensure a meaningful comparison, we excluded non-memory-intensive programs that performed fewer than 10,000 memory allocations per second. In such cases, the performance overhead is predominantly dominated by tag checking, and other components show negligible deviation from the Glibc baseline. The overall result is presented in Figure \ref{ct_buildup}.

\vspace{0.1cm}
\noindent\textbf{Memory Management.} As illustrated in Figure~\ref{ct_buildup}, ClusterTag introduces approximately  6\%-15\% runtime overhead in memory management. This overhead results from ClusterTag's strategy of periodically releasing all free pages to the kernel, which increases the likelihood of page faults during subsequent allocations. Notably, in \textit{623.xalancbmk\_s}, ClusterTag achieves a 13\% performance improvement. This improvement is attributed to the programs custom deallocation function, which performs a linear search, whereas ClusterTag allocates same-type objects from contiguous memory pages, significantly enhancing memory-access locality.

\vspace{0.1cm}
\noindent\textbf{Tag Assignment.} ClusterTag's tag assignment adds just 0.2\%-1.3\% overhead, since all chunks in a cluster can be retagged with a single traversal. However, in \textit{623.xalancbmk\_s}, this overhead increases significantly to 7\%, primarily due to a higher branch misprediction rate. Further analysis reveals that this program contains numerous pointer array search and traversal operations, where storing tags in the high-order bits of pointers affects branch prediction results.

\vspace{0.1cm}
\noindent\textbf{Tag Writing \& Checking.}  The overhead introduced by tag writing ranges from 0.6\% to 1.8\%, primarily due to the need to update shadow memory upon allocation. The program \textit{621.wrf\_s} exhibits notably higher overhead, which is attributed to its frequent allocation of large memory chunks exceeding 0x10000 bytes. For such chunks, ClusterTag writes shadow memory with a 16-to-1 byte mapping, similar to HWASan, resulting in significant memory operations. Tag checking accounts for the most significant portion of ClusterTag's runtime overhead, with an average slowdown of 124\%. This considerable performance cost is intrinsic to our software-based approach, in which each memory access requires additional checking instructions. We discuss the potential integration of ClusterTag with MTE hardware in Section~6.1, which may leverage hardware acceleration to substantially mitigate this overhead.

\FloatBarrier
\subsubsection{SPEC CPU 2017 Memory Overhead}\quad \\
Table \ref{cluster2017memory} compares the memory overhead between the original program (based on Glibc) and programs hardened with HWASan and ClusterTag. Experimental results indicate that both HWASan and ClusterTag incur approximately a 7.4\% increase in memory overhead compared to the original program \footnote{$(3,104-2,889)/2,889 \approx 7.4\%$}. This overhead primarily stems from the additional space allocated for shadow memory. Specifically, since both methods utilize an 8-bit tag encoding scheme, every 16 bytes of application memory corresponds to 1 byte of tag memory, theoretically resulting in roughly a 6\% increase in memory usage.

\begin{table}[h]
    \centering
    \caption{Memory usage of ClusterTag on SPEC CPU 2017 (HW=HWASan, CT=ClusterTag, PTE=Page Table Overhead)}
    \label{cluster2017memory}
    
    \renewcommand{\arraystretch}{1.15}
    \small
    \setlength{\tabcolsep}{2pt}
    
    \begin{tabular}{l r r | r r r | r r}
    \toprule
    \multirow{2}{*}{\textbf{Benchmark}} & \textbf{Glibc} & \textbf{HW} & \multicolumn{3}{c|}{\textbf{ClusterTag}} & \multicolumn{2}{c}{\textbf{PTE} (MB)} \\
    \cmidrule(lr){4-6} \cmidrule(lr){7-8} 
    & (MB) & (MB) & $(d=5)$ & $(d=10)$ & $(d=20)$ & HW & CT \\
    \midrule
    600.perlbench\_s  &     143 &     172 &     177 &     178 &     177 &  1.24 &  6.74 \\
    605.mcf\_s        &   3,598 &   3,941 &   3,932 &   3,941 &   3,942 &   8.0 &  9.07 \\
    607.cactuBSSN\_s  &   6,684 &   7,110 &   7,097 &   7,093 &   7,093 & 14.54 &  15.1 \\
    619.lbm\_s        &   3,227 &   3,431 &   3,429 &   3,428 &   3,429 &   7.2 &  7.26 \\
    621.wrf\_s        &     179 &     189 &     207 &     206 &     207 &  1.16 &  1.89 \\
    623.xalancbmk\_s  &     406 &     492 &     512 &     513 &     520 &  1.93 & 12.09 \\
    628.pop2\_s       &   1,433 &   1,488 &   1,494 &   1,494 &   1,494 &  3.81 &  5.88 \\
    631.deepsjeng\_s  &   6,881 &   7,310 &   7,302 &   7,304 &   7,302 & 14.48 & 14.43 \\
    638.imagick\_s    &   4,234 &   4,506 &   4,503 &   4,500 &   4,502 &  9.87 &  9.54 \\
    641.leela\_s      &      16 &      24 &      20 &      20 &      20 &  0.69 &  8.61 \\
    644.nab\_s        &     562 &     682 &     688 &     688 &     688 &  2.13 &  5.85 \\
    657.xz\_s         &   7,307 &   7,901 &   7,731 &   7,780 &   7,715 & 16.21 & 15.75 \\
    \midrule
    \textit{Average}   &   2,889 &   3,104 & \cellcolor[gray]{0.9}   3,091 & \cellcolor[gray]{0.9}   3,095 & \cellcolor[gray]{0.9}   3,091 & 6.77  & 9.35 \\
    \bottomrule
    \end{tabular}
\end{table}

Unlike HWASan's contiguous memory allocation, ClusterTag randomly distributes memory objects across the address space. Experimental results show that this randomized memory layout achieves slight memory savings (within approximately 0.2\%) across various randomization densities (5, 10, and 20). This advantage is primarily due to its cluster-based randomization unit: although memory objects are randomly distributed from a global perspective, they are stored contiguously within a local cluster, occupying entire physical pages. To prevent excessive page table overhead, ClusterTag further employs the \textit{pool} mechanism to constrain the randomization range of individual clusters. As shown in Table \ref{cluster2017memory}, randomized processes incur an average increase of only \textasciitilde2MB in page table entries, negligible compared to the average process memory consumption (\textasciitilde4GB). We observe that \textit{623.xalancbmk\_s} and \textit{641.leela\_s} incur higher PTE overhead than other programs, as their sparse \textit{region} distributions within the \textit{pool} area require extensive PTEs to cover the scattered address space.

\FloatBarrier
\subsection{Security Evaluation}
To answer research question \textbf{RQ2}, we empirically evaluate ClusterTag against existing tag assignment strategies, measuring its effectiveness in vulnerability detection and resistance to tag probing.

\begin{table*}[t]
    \centering
    \caption{Comparison of ClusterTag with other tag assignment strategies on vulnerability detection results}
    \label{ct_juliet}

    \small
    \setlength{\tabcolsep}{3pt}

    \begin{tabular}{l c | cc >{\columncolor[gray]{0.9}}c | cc >{\columncolor[gray]{0.9}}c | cc >{\columncolor[gray]{0.9}}c | cc >{\columncolor[gray]{0.9}}c | cc >{\columncolor[gray]{0.9}}c}
      \toprule
      \multirow{2}{*}{\textbf{Benchmark}} & 
      \multirow{2}{*}{\makecell{All\\Cases}} & 
      \multicolumn{3}{c|}{ClusterTag (d=5)} & 
      \multicolumn{3}{c|}{HWASan} & 
      \multicolumn{3}{c|}{Scudo} & 
      \multicolumn{3}{c|}{PartitionAlloc} & 
      \multicolumn{3}{c}{StickyTags} \\
      \cmidrule(lr){3-5} \cmidrule(lr){6-8} \cmidrule(lr){9-11} \cmidrule(lr){12-14} \cmidrule(lr){15-17}
      & & TP & FN & PN & TP & FN & PN & TP & FN & PN & TP & FN & PN & TP & FN & PN\\
      \midrule
      Heap Buffer Overflow (122)  & 3,342 & 3,306  &  36   & \textbf{0}     & 3,227 &  36 & \textbf{79}    & 2,592 & 654   & \textbf{96}  & 0     & 1,170  & \textbf{2,172} & 2,592 & 654   & \textbf{96} \\
      Buffer Underwrite (124)     &   976 &    480 & 496   & \textbf{0}     &    84 & 496 & \textbf{396}   & 480   & 496   & \textbf{0}   & 0     & 976    & \textbf{0}     & 448   & 480   & \textbf{48} \\
      Buffer Overread (126)       &   678 &    336 & 342   & \textbf{0}     &   295 & 342 & \textbf{41}    & 288   & 342   & \textbf{48}  & 0     & 342    & \textbf{336}   & 528   & 108   & \textbf{48} \\
      Buffer Underread (127)      &   976 &    384 & 592   & \textbf{0}     &    46 & 592 & \textbf{338}   & 480   & 496   & \textbf{0}   & 0     & 976    & \textbf{0}     & 448   & 480   & \textbf{48} \\
      Double Free (415)           &   816 &    816 &   0   & \textbf{0}     &   651 &   0 & \textbf{165}   & 0     & 816   & \textbf{0}   & 816   & 0      & \textbf{0}     & 0     & 816   & \textbf{0} \\
      Use After Free (416)        &   394 &    330 &  64   & \textbf{0}     &   253 &  64 & \textbf{77}    & 284   & 0     & \textbf{110} & 306   & 0      & \textbf{88}    & 0     & 394   & \textbf{0} \\
      \midrule
      \textbf{Juliet Summary}     & 7,182 & 5,652  & 1,530 & \textbf{0}     & 4,556 & 1,530 & \textbf{1,096} & 4,124 & 2,804 & \textbf{254} & 1,122 & 3,464  & \textbf{2,596} & 4,016 & 2,932 & \textbf{240} \\
      \bottomrule
    \end{tabular}
\end{table*}

\subsubsection{Vulnerability Dataset}\quad \\
Our security evaluation is conducted based on the Juliet C/C++ \cite{juliet} vulnerability dataset. It provides thousands of vulnerable code examples covering 118 CWE (Common Weakness Enumeration) categories, including buffer overflows and use-after-free vulnerabilities, which helps evaluate ClusterTag's capability boundaries in various vulnerability detection tasks. Notably, the Juliet dataset is primarily designed for evaluating static analysis tools and exhibits significant limitations for dynamic detection tool evaluation. Due to its lack of runtime state diversity, out-of-bounds pointers and dangling pointers always point to unallocated regions, making them easily detectable by all tag-based sanitizers. To comprehensively evaluate the differences in tag collision probability between ClusterTag and other approaches, we extended the Juliet dataset and detection tools in the following three aspects:

\begin{itemize}
    \item \textbf{Memory State Diversification.} To disrupt the homogeneous memory states across Juliet's test cases, we instrument all memory allocation/deallocation operations to randomly execute 5,000 memory operations, including randomly allocating objects of different sizes and freeing them. This diversification helps create more realistic memory states and challenges the detection tools.
    
    \item \textbf{Vulnerability Triggering Constraints.} Some vulnerable programs in the Juliet dataset are triggered probabilistically through random branches or user input. To ensure consistent evaluation across sanitizer tools, our experiment constrains these behaviors to make vulnerability triggering deterministic.
    
    \item \textbf{Intrinsic Memory Function Instrumentation.} Unlike hardware-assisted sanitizers (StickyTags, Scudo, PartitionAlloc), HWASan relies on software instrumentation and thus requires special handling of memory intrinsic functions. We expanded HWASan by adding instrumentation for eight more functions (\textit{strcat}, \textit{strncat}, \textit{strcpy}, \textit{strncpy}, \textit{strncmp}, \textit{strncasecmp}, \textit{snprintf}, \textit{wcscpy}) beyond the original three (\textit{memmove}, \textit{memcpy}, \textit{memset}). ClusterTag also monitors these memory intrinsic functions to maintain comparable capabilities with HWASan.
\end{itemize}

\subsubsection{Juliet Results}\quad \\
We selected six CWE categories from the Juliet dataset, focusing on spatial and temporal memory safety issues. By conditionally compiling the vulnerable branches, we generated 7,182 vulnerable programs as test cases. To comprehensively examine tag collision probability differences between ClusterTag and other sanitizers, each test case is executed for 500 iterations. The evaluation results are quantified using three metrics: True Positive (\textcolor{mygreen}{\textbf{TP}}) denotes that the vulnerability is detected in all iterations with deterministic behavior. False Negative (\textcolor{myred}{\textbf{FN}}) refers to cases where the vulnerability is consistently missed across all executions, implying a complete failure of detection. Probabilistic Negative (\textcolor{blue}{\textbf{PN}}) represents cases where the vulnerability is detected in some iterations but missed in others, indicating potential tag collisions. We use miss rate to denote the proportion of false negatives across all rounds.

\vspace{0.1cm}
\noindent\textbf{ClusterTag.} As shown in Table \ref{ct_juliet}, ClusterTag provides higher probabilistic memory protection, it consistently detected 5,672 out of 7,182 vulnerable code segments across 500 repeated tests (\textcolor{mygreen}{\textbf{TP}}). In this evaluation, the randomization density parameter $d$ was set to 5. It can be further increased to 10 or even 20, which would provide an even lower tag collision probability while maintaining a comparable performance overhead. To understand the remaining undetected cases, we manually analyzed the 1,530 false negatives, and identified two main causes (\textcolor{myred}{\textbf{FN}}). The first is missing tag assignment at the source, such as stack-allocated objects that are not instrumented. The second is missing tag checks at the sink, including standard library functions (e.g., printf) and intra-structure overflows that bypass field-level detection. These findings reveal current limitations in ClusterTag's coverage and indicate future directions for improvement.

\vspace{0.1cm}
\noindent\textbf{HWASan (\textit{with Random Tag Assignment}).} As shown in Table \ref{ct_juliet}, our analysis reveals significant detection gaps between HWASan and ClusterTag. While both tools failed to detect 1,530 vulnerabilities (\textcolor{myred}{\textbf{FN}}), HWASan additionally exhibited probabilistic negatives on 1,096 out of the 5,652 vulnerabilities successfully detected by ClusterTag (\textcolor{blue}{\textbf{PN, miss rate: 0.4\%}}). These false negatives stem from three types of tag collision issues across specific vulnerability classes. For adjacent overflows in CWE-122 to CWE-127, HWASan's separate chunk header design allows ``off-by-one'' out-of-bounds accesses to reach adjacent memory objects, resulting in false negatives when tags match. For non-adjacent overflows, random illegal memory accesses may inadvertently target memory objects with identical tags, thus bypassing HWASan's security checks. For temporal memory errors (CWE-415 and CWE-416), HWASan's reuse of tags during object reallocation causes these errors to go undetected when freed objects are reallocated with the same tag as the original allocation.
\par
It is further observed that HWASan achieves a higher deterministic detection (\textcolor{mygreen}{\textbf{TP}}) on CWE-122 than hardware-based solutions. This is due to HWASan's tail-tag mechanism, which precisely records the requested allocation size within each memory slot. When an object does not occupy the entire slot, the unused padding bytes serve as a guard region, enabling deterministic detection of contiguous buffer overflows.

\vspace{0.1cm}
\noindent\textbf{Scudo (\textit{with Staggered Tag Assignment}).} Scudo divides 16 tags into two groups based on memory address parity (odd or even). This design detects all adjacent overflows in CWE-122 to CWE-127 (\textcolor{mygreen}{\textbf{TP}}). However, non-adjacent overflows can bypass adjacent blocks and access memory regions with the same parity, causing probabilistic misses in CWE-122 and CWE-126  (\textcolor{blue}{\textbf{PN, miss rate: 4\%}}). Additionally, Scudo deterministically failed to detect 1,988 vulnerabilities, due in part to its lack of support for stack variable protection and also because invalid pointers target padding bytes in unaligned memory blocks.
\par
For temporal vulnerabilities, Scudo assigns different tags for reallocated chunks, enabling deterministic detection of 284 use-after-free cases in CWE-416 (\textcolor{mygreen}{\textbf{TP}}). However, since Scudo maintains historical tags for only one reallocation cycle, it remains susceptible to probabilistic tag collisions in vulnerabilities involving multiple reallocation cycles (\textcolor{blue}{\textbf{PN, miss rate: 13.6\%}}). For double-free cases in CWE-415, Scudo uses the chunk's status bit to detect errors instead of relying on tag verification, leading to deterministic misses (\textcolor{myred}{\textbf{FN}}).

\vspace{0.1cm}
\noindent\textbf{PartitionAlloc (\textit{with Fixed Tag Assignment}).} PartitionAlloc adopts a random tag assignment strategy similar to HWASan, which results in probabilistic misses for 2,508 cases within CWE-122 to CWE-127 (\textcolor{blue}{\textbf{PN, miss rate: 7\%}}). Additionally, both PartitionAlloc and StickyTags divide memory into small and large chunks based on allocation size, with only small chunks tagged via MTE. Because the two regions are not properly isolated, an overflow from a small chunk can directly reach an adjacent large chunk, as illustrated in Listing \ref{partitioncase}. This design allows tagged pointers to dereference untagged addresses under MTE, causing deterministic misses (\textcolor{myred}{\textbf{FN}}).
\par
Regarding temporal behavior, PartitionAlloc employs a fixed tag assignment scheme where tags increment by one after each deallocation and wrap around after 15 reuse cycles. This strategy significantly extends the time window between identical tags, enabling deterministic detection of 1,122 cases in CWE-415 and CWE-416 (\textcolor{mygreen}{\textbf{TP}}). However, probabilistic misses occur in 88 cases of CWE-416, corresponding to exactly 15 reallocations after the first deallocation(\textcolor{blue}{\textbf{PN, miss rate: 3.8\%}}). In contrast, ClusterTag fully releases old clusters and reallocates memory at randomized addresses (see Section 4.3.2), thus achieving deterministic detection for these cases.

\begin{lstlisting}[caption={FN example in PartitionAlloc and StickyTags}, label={partitioncase}]
#include <stdio.h>

void main() {
    char* large_chunk = malloc(0x10000);
    char* small_chunk = malloc(100);
    // flaw. Buffer Underwrite
    small_chunk[-8] = 'c';
}
\end{lstlisting}

\vspace{0.1cm}
\noindent\textbf{StickyTags (\textit{with Fixed Tag Assignment}).} StickyTags adopts a fixed tag assignment scheme, similar to PartitionAlloc, with tags increasing sequentially by one across adjacent memory blocks and cycling every 16 allocations. This strategy provides each allocation with a protection zone of up to $15 \times ChunkSize$, theoretically enabling deterministic detection of all adjacent overflows in CWE-122 to CWE-127 (\textcolor{mygreen}{\textbf{TP}}). However, due to insufficient isolation between small and large memory chunks, 1,722 cases in these categories remain undetected (\textcolor{myred}{\textbf{FN}}). For non-adjacent overflows, the limited protection range (e.g., a 0x20-byte object only has a 480-byte buffer zone) is inadequate to cover the $\pm$1,024-byte random access pattern in the Juliet dataset, leading to 240 probabilistic misses (\textcolor{blue}{\textbf{PN, miss rate: 6.3\%}}).
\par
For temporal vulnerabilities, StickyTags retains the same tag for each memory address to reduce performance overhead from tag reassignment. This optimization, however, prevents it from detecting any temporal memory errors in CWE-415 and CWE-416 (\textcolor{myred}{\textbf{FN}}).

\FloatBarrier
\subsubsection{Defending Against Tag Probing Attacks}\quad \\
Recent research \cite{stickytag,kim2024tiktag} has revealed that side-channel vulnerabilities in MTE allow attackers to perform tag probing without triggering alerts. In this section, we will demonstrate how ClusterTag defends against such attacks by increasing the tag collision threshold and enhancing the diversity of memory states.

\begin{figure}[h]
  \centering
\setlength{\leftskip}{10pt}
  \includegraphics[width=3in]{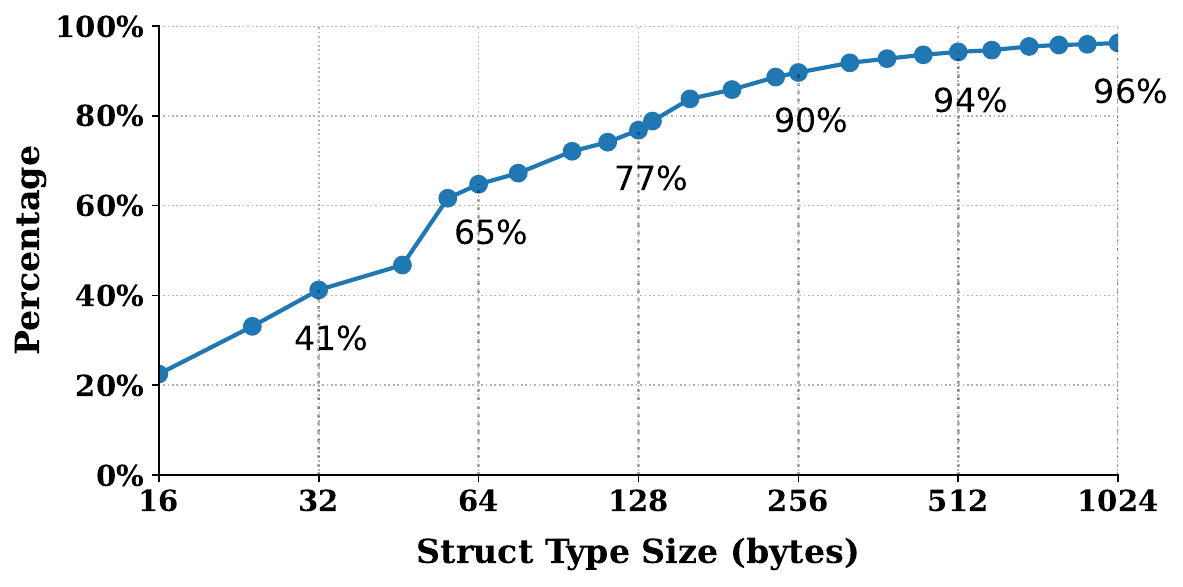}
\caption{Distribution of struct type sizes in SPEC CPU 2017.}

\label{ct_struct}
\end{figure}

\vspace{0.1cm}
\noindent\textbf{Increasing the Collision Distance Threshold.} ClusterTag defends against tag probing attacks by separating the overlap between tag collision ranges and impact ranges of illegal pointers. This design ensures that even if attackers infer the tag of a target address through side channels, they cannot exploit illegal pointers to launch effective attacks. Taking type confusion, a typical non-adjacent overflow, as an example, we conducted a statistical analysis of all structure sizes in the SPEC CPU 2017 benchmark suite. As shown in Figure \ref{ct_struct}, 96\% of structures in SPEC CPU 2017 have sizes not exceeding 1,024 bytes (compared to 99\% in SPEC CPU 2006 \cite{stickytag}). This result indicates that the effective range of illegal pointers is typically limited to $\pm$1KB, which is significantly smaller than the 8,192-byte minimal collision distance provided by ClusterTag. Furthermore, we analyzed the integer overflow vulnerability CVE-2018-14883 in the Magma dataset. Among all 594 PoCs, the illegal pointer impact range was between -2,095 and 1,791 bytes, well below ClusterTag's 8,192-byte collision distance threshold.
\par
Complementing this spatial defense, ClusterTag also applies temporal protection by employing quarantine tagging (Section 4.4), which ensures no tag collisions occur within 16 reallocation rounds, thereby providing deterministic protection against short-term dangling pointers similar to secure heap allocators like Scudo \cite{scudo}. To mitigate prolonged probing attempts, ClusterTag reclaims idle heap clusters and relocates subsequent allocations to new randomized addresses (Section 4.3.2). Our analysis of SPEC CPU 2017 shows that heap clusters are typically reclaimed after an average of 275 reallocation rounds, effectively invalidating addresses referenced by dangling pointers within a bounded timeframe and substantially limiting attackers' ability to conduct sustained tag probing.

\vspace{0.1cm}
\noindent\textbf{Increasing Heap State Diversity.} To enhance the effectiveness of tag probing attacks, attackers often employ heap-spraying techniques to construct predictable heap layouts, thereby increasing the likelihood of triggering tag collisions or bypassing quarantine mechanisms. ClusterTag mitigates such advanced attacks by diversifying heap states along both spatial and temporal dimensions. In the spatial dimension, ClusterTag assigns unique tags within each cluster and randomizes inter-cluster distances, thereby increasing spatial entropy and reducing the predictability necessary for successful spraying. In the temporal dimension, it randomly selects clusters and uses a circular shift to rotate tags for reused memory chunks. This design forces attackers to precisely control the lifetimes of all 256 chunks in a cluster and repeatedly reallocate memory for at least 16 rounds to reintroduce a specific tag. Given the inherent noise and nondeterminism of real-world runtime environments, achieving such precision is highly impractical, making sustained tag probing attacks extremely difficult to carry out.

\FloatBarrier
\subsection{Quantification of Tag Collision Distance}
In Section 2.3, we proposed three metrics for measuring tag collision: minimum distance, average distance, and unpredictability, and analyzed existing tag assignment strategies accordingly. To answer the research question \textbf{RQ3}, we further quantify ClusterTag using the aforementioned metrics, demonstrating its advantages in vulnerability detection and resistance to tag probing attacks.

\begin{table}[b]
    \centering
    \caption{Spatial (S, chunk size units) and Temporal (T, allocation round units) tag collision distance between ClusterTag and HWASan}  
    \label{ct_quantify}
\setlength{\tabcolsep}{4pt}
    \renewcommand{\arraystretch}{1}  

    \begin{tabular}{c cc  cc  cc}
        \toprule
\multirow{2}{*}{\textbf{Strategy}} & \multicolumn{2}{c}{\textbf{Minimum}} & \multicolumn{2}{c}{\textbf{Average}} & \multicolumn{2}{c}{\textbf{Unpredictability}} \\
        \cmidrule(lr){2-3} \cmidrule(lr){4-5} \cmidrule(lr){6-7}
         & \textit{S} & \textit{T} & \textit{S} & \textit{T} & \textit{S (bit)} & \textit{T (bit)} \\

        \midrule
        
        \textit{ClusterTag} & \cellcolor[gray]{0.9}256 & \cellcolor[gray]{0.9}16 & \cellcolor[gray]{0.9}$256 \cdot d$ & 510 & \cellcolor[gray]{0.9}$8.7+H(G(\frac{1}{d}))$ & 9.53 \rule[-6pt]{0pt}{20pt} \\
        \textit{HWASan} & 1 & 1 & 256 & \cellcolor[gray]{0.9}543 & $H(G(\frac{1}{256}))$ & \cellcolor[gray]{0.9}10.53 \rule[-6pt]{0pt}{20pt} \\
        \bottomrule
    \end{tabular}
\end{table}

\begin{figure*}[t]
    \centering
    \includegraphics[width=6.3in]{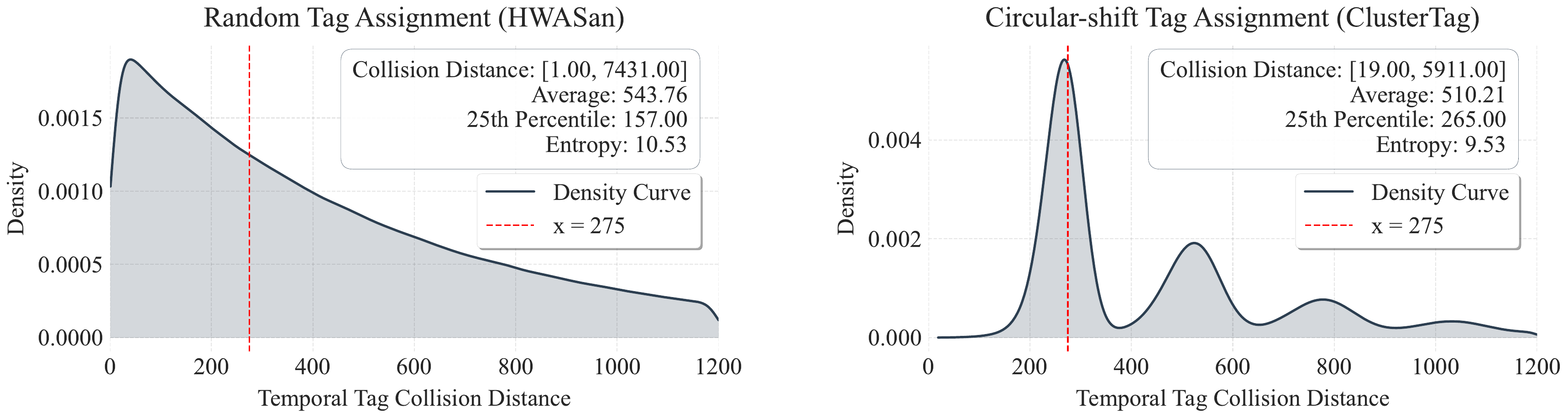}
    \caption{Temporal tag collision distance distribution}
    \label{ct_distribution}
\end{figure*}

\par
We select the random tag assignment strategy as the primary comparison baseline, as this strategy is not only directly adopted by HWASan \cite{hwasan} and Ptmalloc \cite{ptmalloc}, but also serves as the design foundation for other assignment strategies. For example, Scudo \cite{scudo} uses different tag pools for odd and even memory chunks and randomly selects tags within each pool; PartitionAlloc \cite{partition} also adopts random assignment in the spatial dimension. Finally, this section does not analyze fixed assignment strategies (StickyTags \cite{stickytag} and PartitionAlloc's temporal dimension strategy) as they are straightforward: they always maximize both minimum and average collision distances but completely sacrifice unpredictability.

\subsubsection{Spatial Collision Distance Analysis}\quad \\
The columns labeled ``S'' in Table \ref{ct_quantify} show three types of objectives for spatial collision distances. The units for the first two categories are in chunk size, while the third category is in bits.
\par

\vspace{0.1cm}
\noindent\textbf{Minimum.} ClusterTag introduces at least one cluster-sized gap between clusters, thus its minimum collision distance is $256 \cdot ChunkSize$, as shown in Table \ref{ct_quantify}. This means that even for the smallest aligned memory objects (\textit{0x20} bytes) in ClusterTag, there is no other object with the same tag within $\pm$8,192 bytes. By comparison, HWASan independently assigns random tags to each chunk, resulting in a minimum collision distance of only $1 \cdot ChunkSize$. StickyTags provides incremental tags for each adjacent memory block, with a minimum collision distance of $16 \cdot ChunkSize$.

\vspace{0.1cm}
\noindent\textbf{Average.} When tags are selected with equal probability from a 256-value space, HWASan's spatial collision distance approximately follows a geometric distribution $P(X = k) = (1-p)^{k-1} \cdot p$. We denote this distribution as $G(p)$, where $p=\frac{1}{256}$. By comparison, ClusterTag's spatial collision distance is more complex, considering two identical tags appearing in different clusters. Let $D$ denote the number of clusters separating them, and $Z_1, Z_2$ represent the relative positions of these two tags within their respective clusters. The spatial collision distance can then be expressed as $256 \cdot D + (Z_2-Z_1)$. In this distance model, the inter-cluster distance $D$ follows a geometric distribution $G(\frac{1}{d})$ where $d$ is the density parameter; $Z_1$ and $Z_2$ follow uniform distributions; $Z_2-Z_1$ follows a triangular distribution.
\par
Based on the above distributions, it can be calculated that HWASan's average collision distance is $E(X)=\frac{1}{p}=256$. ClusterTag's average collision distance is $E(256 \cdot D + (Z_2-Z_1))=256 \cdot E(D) + E(Z_2-Z_1)=256 \cdot d + 0=256 \cdot d$.

\vspace{0.1cm}
\noindent\textbf{Unpredictability.} We use entropy to evaluate the unpredictability of tag collision distances. For HWASan with an 8-bit tag encoding, substituting its geometric distribution $G(\frac{1}{256})$ into the entropy formula yields: $H(G(p)) = -\sum_{k=1}^{\infty} (1-p)^{k-1} \cdot p \cdot \log_2 [(1-p)^{k-1} \cdot p] \approx 9.44$ bits. In comparison, ClusterTag employs heap randomization techniques to surpass the inherent entropy limitations associated with tag encoding, thus achieving higher and adjustable entropy values. Specifically, ClusterTag's collision distance comprises two independent events: cluster selection and tag selection. Given this independence, we can approximate the upper bound of ClusterTag's entropy using the $H(256 \cdot D)+H(Z_2-Z_1)$ \footnote{$H(X+Y) \le H(X,Y) = H(X)+H(Y)$, when $X$ and $Y$ are independent events}, where calculations indicate $H(256 \cdot D)=H(G(\frac{1}{d}))$ and $H(Z_2-Z_1) \approx 8.72$ bits (entropy of triangular distribution).
\par
Based on the above distributions, when ClusterTag's randomization density $d$ is set to 5, 10, and 20, the achievable entropy values are 12.33, 13.41, and 14.45 bits respectively, all exceeding the 9.44 bits provided by HWASan.

\subsubsection{Temporal Collision Distance Analysis}\quad \\
The temporal tag states involve random factors such as memory allocation and deallocation, making precise mathematical modeling challenging. As an alternative, we used Monte Carlo simulation to plot tag collision distance distributions for both HWASan and ClusterTag, as shown in Figure \ref{ct_distribution}. Our simulation examined 256 chunks in a single cluster. In each round, we randomly selected 1-240 chunks and assigned new tags using either random generation (HWASan) or circular shift (ClusterTag). We then calculated the round interval between identical tags for each chunk and recorded these as samples. The distribution was generated from over 2 million samples using Gaussian kernel density estimation.

\vspace{0.1cm}
\noindent\textbf{Minimum.} As shown in Table \ref{ct_quantify}, the minimum collision distances for the random selection strategy (HWASan) and the circular right-shift strategy (ClusterTag) are 1 and 16, respectively. The advantage of ClusterTag is attributed to the reservation of a set of tags (16 in our experiments) for each cluster. These reserved tags effectively extend the minimum chain length in the circular shift, providing a longer cycle period before chunks are reassigned to the same tag. It should be noted that the probability of ClusterTag occurring at the minimum distance is very low, and it is only possible when a cluster is continuously reused for 16 rounds and the historical tag happens to cycle back to the original chunk. As shown in Figure \ref{ct_distribution}, the observed minimum distance in our simulation is actually 19.

\vspace{0.1cm}
\noindent\textbf{Average.} 
The average collision distances for HWASan and ClusterTag are 543.76 and 510.21, respectively, as shown in Figure \ref{ct_distribution}. Although ClusterTag's distance is slightly lower, the distribution patterns of these two strategies differ significantly. Specifically, the random strategy exhibits a typical skewed distribution, whereas the circular right-shift strategy demonstrates a distinct multi-modal distribution. This fundamental difference results in collision distances being concentrated around shorter distances in the former strategy, while in the latter strategy, they are predominantly around the first peak (approximately 250 rounds). As evidenced by the 25th percentile values in Figure \ref{ct_distribution}, the circular-shift strategy demonstrates a significantly higher value of 265 compared to the random strategy's 157.

\vspace{0.1cm}
\noindent\textbf{Unpredictability.} Although the entropy of ClusterTag (9.53 bits, as shown in Figure \ref{ct_distribution}) is slightly lower than HWASan's 10.53 bits, and the one-position shift might intuitively suggest exploitable patterns, accurately predicting these patterns remains challenging in practice. First, ClusterTag randomly selects clusters for reuse, which increases the unpredictability of internal cluster states over time. Second, clusters are immediately returned to the kernel once all memory objects are freed (experiencing an average of 275 reuse rounds), further shortening the window of opportunity for adversaries to conduct real-time attacks.

\FloatBarrier 
\section{Discussion \& Future Work}
\subsection{From HWASan (8-bit) to MTE (4-bit).}
Although the current prototype of ClusterTag is based on an 8-bit tag length, it can flexibly adapt to hardware mechanisms with shorter tag encoding spaces, such as MTE and LAM, providing a similar level of security protection. However, when the tag length decreases, the number of heap chunks contained in each cluster is reduced, which leads to inefficient utilization of memory pages, resulting in increased memory fragmentation and notable performance overhead. In the following, we demonstrate how this issue can be mitigated through tag group isolation.

\vspace{0.1cm}
\noindent\textbf{Design.} To overcome the limited tag encoding space in hardware mechanisms like MTE, ClusterTag extends traditional tag-based isolation by introducing group-level partitioning within clusters, forming a ``two-dimensional'' tag isolation scheme. Specifically, each cluster is split into multiple tag groups, with unique tags assigned to chunks inside the same group to prevent intra-group collisions. To avoid inter-group collisions, a 4-bit group ID is encoded in the unused high-order bits (60-63) of pointers. Before every memory access, ClusterTag verifies the pointer's group ID matches the expected group, enforcing group boundaries. This design effectively increases the tag space and mitigates memory fragmentation caused by smaller tag lengths. On the other hand, it is also highly compatible with existing MTE mechanisms: 1) group boundary checking only requires simple bitwise operations at the register level, adding negligible performance overhead without extra memory consumption; 2) the group ID is stored only in pointers and does not require extra tag memory space.

\vspace{0.1cm}
\noindent\textbf{Implementation.} We implemented a prototype integrating ClusterTag with MTE, excluding group boundary checking. Compared to the previous design in this paper, the prototype has made the following key adjustments: 1) We reduce the number of chunks in each cluster from 256 to 128, which still ensures cluster alignment with 4 KB pages even at the minimum slot size ($0x20 \times 128 = 4096$ bytes). 2) We divide the 128 chunks into 16 groups with a 4-bit group ID stored in pointer bits 60-63. 3) We assign tags 1-15 to eight chunks within groups, with each chunk receiving a unique tag to prevent spatial tag collisions, while reserving the remaining 7 tags as a quarantine zone for circular rotation to delay temporal tag collisions.
\par
Performance was evaluated on 14 SPEC CPU 2006 C benchmarks (excluding \textit{402.gcc\_s} which crashed due to memory violations) on a Pixel 8 Pro device. The results show that, compared to Scudo, ClusterTag achieves an average reduction of 2.24\% in memory overhead and 0.38\% in runtime overhead, demonstrating its potential for mitigating vulnerability exploitation in production environments. In future work, we plan to further implement group boundary checking for ClusterTag and conduct comprehensive security and performance evaluations. This extension can be implemented either through minor hardware modifications to MTE or via compile-time instrumentation.

\subsection{Secure Heap Allocator}
Existing secure heap allocators employ isolation and randomization techniques to reduce heap vulnerability exploitation, representing an orthogonal research area to memory vulnerability detection discussed in this paper. These allocators primarily pursue two core objectives: protecting the integrity of object metadata (\textit{chunk header}) and preventing interference between memory objects. For metadata protection, approaches such as Scudo \cite{scudo} and the method proposed by Robertson et al. \cite{robertson} embed checksums within object metadata to detect malicious tampering. Similarly, BiBOP-style allocators like DieHarder \cite{dieharder} store metadata separately from memory objects---a design principle also adopted by ClusterTag. For memory object protection, isolation-based secure allocators establish boundaries between memory objects to prevent mutual interference. For example, PartitionAlloc \cite{partition} employs partitioning and guard pages to mitigate spatial overflows, while Scudo \cite{scudo} introduces temporal delays to create gaps before reallocation. Additionally, randomization-based allocators \cite{diehard, dieharder} randomly distribute memory blocks across the address space, substantially increasing the difficulty for attackers to predict memory layout.
\par
Compared to secure heap allocators, the design novelty of ClusterTag lies in applying traditional heap randomization to the vulnerability detection task, effectively overcoming the limitations of both lines of work. On one hand, ClusterTag employs randomization techniques to address the challenge of cross-cluster tag collisions. On the other hand, it utilize tagged memory techniques to extend the size of randomization units (\textit{clusters}), thereby overcoming the memory fragmentation issues faced by traditional heap randomization schemes such as DieHarder \cite{dieharder}.

\section{Conclusion}
Traditional tag-based sanitizers suffer from probabilistic false negatives due to tag collisions. To mitigate this issue, we propose ClusterTag, an innovative approach that integrates tagged memory with heap randomization, effectively overcoming the inherent limitations of tag encoding space. ClusterTag arranges contiguous memory objects into distinct clusters and applies a circular-shift strategy to assign tags within each cluster. It also employs randomization techniques to effectively mitigate inter-cluster tag collisions. Experimental results demonstrate that ClusterTag achieves slightly better performance (\textasciitilde1\%) compared to HWASan across three different randomization densities (5, 10, and 20). Moreover, ClusterTag successfully detects memory violations in all 500 rounds of repeated testing, verifying the effectiveness of its probabilistic protection mechanism.

\begin{acks}
We greatly appreciate the anonymous reviewers and the associate editor for their valuable feedback. This work is partly supported by the National Natural Science Foundation of China (62272351, 62572354, 62172308, 62172144).
\end{acks}

\bibliographystyle{ACM-Reference-Format}
\balance
\bibliography{reference}

\clearpage
\appendix

\section{Artifact Appendix}

\subsection{Abstract}
This artifact contains the source code for ClusterTag and scripts for \textbf{Functional Evaluation}. Specifically, it provides: (a) the full source code of ClusterTag integrated into LLVM 15.0.0; (b) scripts for conducting functional testing using FFmpeg; (c) scripts for security testing using cases from the Juliet Test Suite to compare ClusterTag against HWASan; and (d) detailed instructions for building the environment using Docker and running the evaluations.

\subsection{Artifact Check-List (Meta-Information)}
\begin{itemize}
    \item \textbf{Run-time environment:} Linux
    \item \textbf{Hardware:} ARMv8 or ARMv9 CPU, at least 8GB RAM.
    \item \textbf{Output:} 1) Functional test reports for FFmpeg video conversion results (small.mp4 $\rightarrow$ small.gif) 2) Security test reports for vulnerability detection/missed results by ClusterTag and HWASan.
    \item \textbf{How much disk space required (approximately)?:} 40GB
    \item \textbf{How much time is needed to prepare workflow (approximately)?:} 10-20 minutes to download and import the Docker image.
    \item \textbf{How much time is needed to complete experiments (approximately)?:} 10-20 minutes
    \item \textbf{Publicly available?:} Yes
    \item \textbf{Code licenses (if publicly available)?:} Apache 2.0
    \item \textbf{Archived (provide DOI)?:} Yes
\end{itemize}

\subsection{How to access}
\subsubsection{Source Code}
github.com/Yiruma96/ClusterTag-repo

\subsubsection{Artifact Evaluation}
zenodo.org/records/16838000

\subsection{Description}
\subsubsection{Hardware dependencies}
A standard ARMv8 or ARMv9 machine that supports the Top Byte Ignore (TBI) mechanism.

\subsubsection{Software dependencies}
Docker

\subsection{Installation}
\begin{verbatim}
tar -xf clustertag.tar.gz
docker load -i clustertag-image.tar
sudo docker run -it clustertag
\end{verbatim}
The built ClusterTag will be located at \texttt{/workspace/build} inside the container.

\subsection{Experiment Workflow}
All testing scripts are located in the \texttt{/workspace/evaluation} directory inside the container.
\begin{verbatim}
# Functional Testing
cd /workspace/evaluation/functional_evaluation
./functional_test.sh
\end{verbatim}
\par
\begin{verbatim}
# Security Testing
cd /workspace/evaluation/security_evaluation
./security_test.sh
\end{verbatim}

\subsection{Expected Results}
\begin{itemize}
    \item \textbf{Functional Testing:} The expected output is a series of success messages, culminating in a confirmation that the functional test passed and the output file \texttt{small.gif} was created.
    \item \textbf{Security Testing:} The script will output the detection and miss rates for both ClusterTag and HWASan across four vulnerability types. The expected result is that ClusterTag achieves a 100\% detection rate (0\% missed) for all test cases. In contrast, HWASan is expected to show a small but non-zero miss rate (e.g., \textasciitilde{}0.4\%) due to probabilistic tag collisions, demonstrating the effectiveness of ClusterTag's design.
\end{itemize}

\section{Web Server Runtime Overhead}
Table \ref{ct_server} presents the runtime overhead of ClusterTag compared to HWASan on three types of web servers. The experiment uses ApacheBench to simulate 500 concurrent clients performing a total of 100,000 HTTP requests. The test targets include four fixed-size pages (50KB, 100KB, 200KB, 500KB) and randomly-sized pages (ranging from 10KB to 1MB). All pages are filled with random characters.

\begin{table}[h]
    \centering
    \caption{Runtime overhead of ClusterTag on Web Servers (based on HWASan)}  
    \label{ct_server}
    \small
    \renewcommand{\arraystretch}{1.1}
    \begin{tabular}{l | c c c c c | c}
    \toprule
                      & 50KB     & 100KB    & 200KB    & 500KB    & Random    & \textbf{Average}  \\
    \hline
    \textbf{Nginx}    & -0.27\%  & -0.90\%  & -0.49\%  & -0.27\%  & -1.35\%   & \cellcolor[gray]{0.9}\textbf{-0.66\%}  \\
    \textbf{Apache}   & -4.40\%  & -3.87\%  & -3.83\%  & -3.96\%  & -3.80\%   & \cellcolor[gray]{0.9}\textbf{-3.97\%}  \\
    \textbf{Lighttpd} & 0.05\%   & -0.30\%  & 0.06\%   & 1.10\%   & -1.04\%   & \cellcolor[gray]{0.9}\textbf{-0.03\%}  \\
    \midrule
    \textbf{Average}  & -1.54\%  & -1.69\%  & -1.42\%  & -1.04\%  & -2.06\%   & \cellcolor[gray]{0.9}\textbf{-1.55\%}  \\
    \bottomrule
    \end{tabular}
\end{table}

The evaluation results indicate that ClusterTag achieves better execution efficiency compared to HWASan on Nginx, Apache, and Lighttpd servers (improvements of 0.66\%, 3.97\%, and 0.03\%, respectively). Overall, the average performance improvement across the three servers is 1.55\%, which demonstrates the practicality of ClusterTag's randomization and tagging assignment strategy in highly concurrent, real-world applications. Additionally, the data presented in columns 2-6 of Table \ref{ct_server} illustrate that ClusterTag maintains a similar time overhead to HWASan when handling memory pages of various sizes, further highlighting ClusterTag's stability in managing memory objects of different scales.

\section{Vulnerability Detection Results on Magma}
To evaluate ClusterTag's vulnerability detection capability in real-world applications, we selected four open-source programs from the Magma \cite{magma} dataset for testing and compared ClusterTag with random tag allocation strategies represented by HWASan. The Magma dataset also provides other real-world vulnerable programs. However, we excluded these from our experiments as OpenSSL and SQLite exhibited compilation failures when built with HWASan instrumentation, whereas Poppler failed to compile under default compilation options.
\par
As shown in Table \ref{ct_magma}, Magma provides a total of 25,407 PoCs for \textit{libpng}, \textit{libtiff}, \textit{libxml2}, and \textit{php}, covering 20 CVE vulnerabilities. After executing each PoC repeatedly for 500 rounds, the experimental results indicate that both ClusterTag and HWASan failed to detect 24,164 memory violations, which is consistent with the report from PACMem \cite{pacmem}. The majority of these false negatives occurred because the vulnerability types were beyond the detection scope of these sanitizers, such as null pointer dereferences and uninitialized memory accesses. However, for the remaining 1,243 PoCs, ClusterTag deterministically detected all memory violations across all 500 iterations, whereas HWASan probabilistically missed 594 of these cases.

\begin{table}[h]
    \centering
    \caption{Comparison of ClusterTag and HWASan Vulnerability Detection Results on Magma}  
    \label{ct_magma}
    \setlength{\tabcolsep}{3.5pt}
    \begin{tabular}{l c | cc >{\columncolor[gray]{0.9}}c | cc >{\columncolor[gray]{0.9}}c}
      \toprule
      \multirow{2}{*}{\textbf{Benchmark}} & 
      \multirow{2}{*}{\makecell{All\\Cases}} & 
      \multicolumn{3}{c|}{ClusterTag (d=5)} & 
      \multicolumn{3}{c}{HWASan} \\
      \cmidrule(lr){3-5} \cmidrule(lr){6-8}
      & & TP & FN & PN & TP & FN & PN \\
      \midrule
      libpng                      &   634  &     0 &   634  & \textbf{0} &     0 &   634  & \textbf{0}   \\
      libtiff                     & 3,716  &   115 & 3,601  & \textbf{0} &   115 & 3,601  & \textbf{0}    \\
      libxml2                     & 19,614 &     0 & 19,614 & \textbf{0} &     0 & 19,614 & \textbf{0}    \\
      php                         & 1,443  & 1,128 &   315  & \textbf{0} &   534 &   315  & \textbf{594}  \\
      \midrule
      \textbf{Summary}            & 25,407 & 1,243 & 24,164 & \textbf{0} &   649 & 24,164 & \textbf{594} \\
      \bottomrule
    \end{tabular}
\end{table}

Further manual analysis reveals that all 594 PoCs missed by HWASan originate from CVE-2018-14883 in the \textit{php} application. This vulnerability exploits an integer overflow within the \textit{Exif} module, resulting in non-contiguous out-of-bounds accesses to the \textit{exif\_offset\_info} structure. Analysis of the execution traces from the 594 PoCs demonstrated that these illegal accesses ranged from -2,095 to 1,791 bytes around the structure. Since HWASan allocates heap chunks within contiguous regions and assigns random tags independently for each chunk, out-of-bounds accesses can potentially land in memory objects with identical tags, leading to false negatives. In contrast, ClusterTag randomly disperses clusters throughout the address space and strictly ensures that they are separated by at least 256 chunk sizes. This means that even for memory objects as small as 32 bytes, ClusterTag guarantees detection of arbitrary out-of-bounds accesses within a range of -8,092 to 8,092 bytes ($256 \times 32$).

\end{document}